\documentclass[11pt, a4paper]{article}
\pdfoutput=1

\usepackage{anysize}
\marginsize{3cm}{3cm}{1.5cm}{1.5cm}
\usepackage[T1]{fontenc}
\usepackage{inputenc}

\usepackage{amsfonts}
\usepackage{amssymb}
\usepackage{graphics}
\usepackage{epsfig}
\usepackage{graphicx}
\usepackage{epsfig}
\usepackage{amssymb}
\usepackage{amsfonts}

\usepackage{amsmath}
\usepackage{xcolor}
\usepackage{float}
\usepackage{enumerate}
\usepackage{slashed}
\usepackage{hyperref}
\hypersetup{colorlinks=true,linkcolor=blue,citecolor=blue}

\usepackage{caption}
\usepackage{subcaption}

\numberwithin{equation}{section}

\newcommand {\beq} {\begin{equation}}
\newcommand {\eeq} {\end{equation}}
\newcommand{\bea}{\begin{eqnarray}}
\newcommand{\eea}{\end{eqnarray}}
\newcommand{\bit}{\begin{itemize}}
\newcommand{\eit}{\end{itemize}}
\def\nl{\nonumber \\}

\def\a{\alpha}

\def\b{\beta}

\def\p{\partial}

\def\le{\left(}
\def\ri{\right)}

\def\beq{\begin{equation}}
\def\eeq{\end{equation}}

\begin{document}

\begin{titlepage}

\begin{flushright}

\end{flushright}
\bigskip
\begin{center}
{\LARGE  {\bf
On subregion action complexity in AdS$_3$ and in the BTZ black hole
  \\[2mm] } }
\end{center}
\bigskip
\begin{center}
{\large \bf  Roberto  Auzzi$^{a,b}$},
 {\large \bf Stefano Baiguera$^{c}$},
  {\large \bf Andrea Legramandi$^{c}$}, \\
   {\large \bf Giuseppe Nardelli$^{a,d}$},
    {\large \bf Pratim Roy$^{e}$}  
 {\large \bf and }     {\large \bf Nicol\`o Zenoni$^{a,b,f}$}
\vskip 0.20cm
\end{center}
\vskip 0.20cm 
\begin{center}
$^a${ \it \small  Dipartimento di Matematica e Fisica,  Universit\`a Cattolica
del Sacro Cuore, \\
Via Musei 41, 25121 Brescia, Italy}
\\ \vskip 0.20cm 
$^b${ \it \small{INFN Sezione di Perugia,  Via A. Pascoli, 06123 Perugia, Italy}}
\\ \vskip 0.20cm 
$^c${ \it \small{Universit\`a degli studi di Milano-Bicocca and INFN, 
Sezione di Milano-Bicocca, \\ Piazza
della Scienza 3, 20161, Milano, Italy}}
\\ \vskip 0.20cm 
$^d${ \it \small{TIFPA - INFN, c/o Dipartimento di Fisica, Universit\`a di Trento, \\ 38123 Povo (TN), Italy} }
\\ \vskip 0.20cm 
$^e${ \it \small{ School of Physical Sciences, NISER, Bhubaneshwar, Khurda 752050, India } }
\\ \vskip 0.20cm 
$^f${ \it \small{ 
Instituut voor Theoretische Fysica, KU Leuven, Celestijnenlaan 200D, B-3001 Leuven, Belgium } }
\\ \vskip 0.20cm 
E-mails: roberto.auzzi@unicatt.it, s.baiguera@campus.unimib.it, \\ a.legramandi@campus.unimib.it,
giuseppe.nardelli@unicatt.it, \\ proy@niser.ac.in, nicolo.zenoni@unicatt.it  \end{center}
\vspace{3mm}

\begin{abstract}
We analytically compute subsystem action complexity for a segment 
in the BTZ black hole background up to the finite term, and we find that
it is equal to the sum of a linearly divergent term proportional to the size 
of the subregion and of a term proportional to the entanglement entropy.
This elegant structure does not survive to more complicated geometries:
in the case of a two segments subregion 
in AdS$_3$, complexity has additional finite contributions.
We give analytic results for the mutual action complexity of a two segments subregion.
\end{abstract}

\end{titlepage}

\section{Introduction}

The AdS/CFT correspondence provides a controlled environment 
to investigate the deep relation between quantum information and gravity.
In holography, entanglement entropy is proportional to the area of extremal surfaces \cite{Ryu:2006bv}.
This result provides a more general framework 
to the idea that the black hole entropy is proportional
 to the area of the event horizon \cite{Bekenstein:1973ur}.
 The issue of entanglement entropy in AdS/CFT has been studied in recent 
 years by many authors,  see \cite{Rangamani:2016dms,Headrick:2019eth} for reviews.
 
The desire of understanding the interior of the  black hole horizon
motivates the investigation of less traditional quantum information quantities.
The growth of the Einstein-Rosen bridge continues for a much longer time scale
compared to the thermalization time, where entanglement entropy saturates.
 This motivates the introduction in holography of the new
quantum information concept of computational complexity 
\cite{Susskind:2014rva,Stanford:2014jda,Susskind:2014moa,Susskind:2018pmk}.
Given a set of elementary quantum unitary operations and a reference state,
quantum complexity is heuristically defined as the minimal number of elementary
operations needed to reach a generic state starting  from the reference one.
Therefore  complexity gives a measure of the difficulty in preparing
 a given state starting from a simple reference state.
A nice geometrical formalism which involves geodesics
in the space of unitary evolutions was introduced in \cite{Nielsen1,Nielsen2}.
In recent years, several attempts have been done to define complexity in quantum field theory.
 When considering free field theories, it is possible to regularize the theory by placing it on a lattice, 
 which reduces the computation of complexity to the case of a set of harmonic oscillators
  \cite{Jefferson:2017sdb,Chapman:2017rqy,Hashimoto:2017fga,Chapman:2018hou,Camargo:2018eof}.
It is still challenging to define complexity for interacting field theories.
In $2$ dimensions, an approach involving the Liouville action was proposed in
\cite{Caputa:2017urj,Caputa:2017yrh,Bhattacharyya:2018wym} and another
based  on Virasoro algebra was studied in \cite{Caputa:2018kdj}.
Another approach, which uses geodesics in the space of sources to define complexity,
was investigated in \cite{Belin:2018bpg}.

A few proposals have been suggested for the holographic dual of complexity:
\begin{itemize}
\item complexity=volume (CV)  \cite{Susskind:2014rva,Stanford:2014jda,Susskind:2014moa}
 relates complexity to the volume $V$ the extremal surfaces anchored at the boundary
   \beq
\mathcal{C}_V = \, {\rm Max}  \le \frac{V}{G L} \ri \, ,
\eeq
where $G$ is the Newton constant and $L$ the AdS length.
\item  complexity=action (CA) \cite{Brown:2015bva,Brown:2015lvg}
 relates it to the action evaluated on the Wheeler-De Witt (WDW) patch, which is the domain of dependence
 of the volume extremal surface 
 \beq
\mathcal{C}_A=  \frac{I_{WDW}}{\pi } \, .
\eeq
\item complexity=spacetime volume  (CV 2.0) \cite{Couch:2016exn}
 links complexity with the spacetime volume $\hat{V}$ of the WDW patch
  \beq
\mathcal{C}_{V 2.0}=  \frac{\hat{V}}{G L^2 } \, .
\eeq
\end{itemize}
Holographic complexity has been recently studied by many groups
in  various asymptotically AdS gravity backgrounds, see for example
\cite{Lehner:2016vdi,Cai:2016xho,Chapman:2016hwi,Carmi:2017jqz,Chapman:2018bqj,
Moosa:2017yvt,Moosa:2017yiz,Chapman:2018dem,Chapman:2018lsv,Barbon:2015ria,Bolognesi:2018ion,
Flory:2018akz,Flory:2019kah,Alishahiha:2017hwg,Akhavan:2018wla}.
The study of holographic complexity can be generalized
also to spacetimes with other UV asymptotics, such as
 Lifshitz theories  and Warped AdS black holes, see e.g.
 \cite{Alishahiha:2018tep,Ghodrati:2017roz,Auzzi:2018zdu,Auzzi:2018pbc,Dimov:2019fxp}.

An interesting  extension of the holographic complexity conjecture
is to consider restrictions to subregions of the boundary conformal field theory.
This is physically motivated by analogy with the entanglement entropy.
Each of the holographic complexity conjectures has a natural 
subregion generalization:
\begin{itemize}
\item the subregion CV  \cite{Alishahiha:2015rta} proposes that the complexity associated to a boundary region $A$
is proportional to the volume of the extremal spatial volume bounded by $A$
and by its Hubeny-Rangamani-Takayanagi (HRT) surface \cite{Hubeny:2007xt}.
\item subregion CA \cite{Carmi:2016wjl} (or CV 2.0) proposes that the subregion complexity
is given by the action (or   the spacetime volume, respectively) of the intersection between
the WDW patch and the entanglement wedge \cite{Headrick:2014cta}.
\end{itemize}
Subregion complexity has been recently 
studied by many authors, e.g.
\cite{Ben-Ami:2016qex,Abt:2017pmf,Abt:2018ywl,Agon:2018zso,
Alishahiha:2018lfv,Caceres:2018blh,Roy:2017kha,Roy:2017uar,Bakhshaei:2017qud,
Bhattacharya:2019zkb,Auzzi:2019fnp,Chen:2018mcc,Auzzi:2019mah}.  
A few options for the quantum information dual 
of holographic subregion complexity have been proposed,
 such as purification or basis complexity \cite{Agon:2018zso}.
 In order to identify the correct quantum field theory dual,
 it is necessary to compute subregion complexity in many
 physical situations. 

In this paper we study the CA and CV 2.0 conjectures for subregions in 
asymptotically AdS$_3$ spacetime.
We  find the following analytic  result for the subregion complexity of a segment 
of length $l$ in the BTZ
\cite{Banados:1992wn,Banados:1992gq} black hole background:
\beq
\label{CABTZ}
 \mathcal{C}_{A}^{\rm BTZ} = \frac{l}{\varepsilon} \frac{c}{6 \pi^2} \log \left(\frac{\tilde{L}}{L} \right)
- \log  \left(\frac{2\tilde{L}}{L} \right) \frac{S^{\rm BTZ}}{\pi^2} + \frac{1}{24} c \, ,
\eeq
where $\tilde{L}$ is a free scale of the counterterm in the action \cite{Lehner:2016vdi}, $\varepsilon$
is the UV cutoff, $c$ the CFT central charge and $S^{\rm BTZ}$ the Ryu-Takayanagi (RT)
entanglement entropy of the segment subregion. 
Equation  (\ref{CABTZ}) is  also valid for the particular case of  AdS$_3$,
which was previously studied in \cite{Carmi:2016wjl,Chapman:2018bqj}.
We find a  similar expression also for the CV 2.0 conjecture, see eq. (\ref{total-btz-bulk}).

The  compact expression (\ref{CABTZ})  follows from surprising 
cancellations in the lengthy direct calculation that we performed.
In particular, the log divergence and part of the finite contribution sum up to
reproduce a term proportional to the entanglement entropy.
This fact does not look as a coincidence and calls for a 
physical interpretation.
 Note that both the coefficients of the linear and log divergent terms
 in  (\ref{CABTZ})  depend on the counterterm scale $\tilde{L}$. 
 We should choose $\tilde{L}>L$ in order to have a positive-definite complexity; 
 consequently, the coefficient of the entropy term in eq.  (\ref{CABTZ})  must be negative.
 Apart from this restriction, $\tilde{L}$ is not specified and its physical
 meaning is still obscure; it might be related to the details of the regularization procedure
 that must be implemented to define complexity on the quantum field theory side.
 We leave these topics for further investigation.

One may wonder if such a simple connection between subregion complexity and 
entanglement entropy is valid also for more general subsystems.
For this reason, we compute action complexity in the case of a 
two  segments subregion in AdS$_3$. This quantity has as before
a linear divergence proportional to the total size of the region and a log
divergence proportional to the divergent part of the entropy.
However, if the separation between the two disjoint segments is small,
there is no straightforward relation between the finite part of complexity and entropy,
see eq. (\ref{CA2}).

The paper is organised as follows. In section \ref{sect:AdS} we review the subregion complexity
calculation for a segment in  AdS$_{3}$. In section \ref{sect:BTZ} we compute the subregion 
complexity for a segment in the BTZ background, and we show that it is related to the entanglement entropy.
In section \ref{sect:AdS-2seg} we calculate subregion complexity for two disjoint segments in AdS$_3$.
In section \ref{sec:mutual}  we discuss mutual complexity. 
In appendix \ref{app:other-reg} we compute the single segment subregion
 complexity in the BTZ background with a different regularization.

{\bf Note added:} While we were finalizing the writing of this paper, Ref. \cite{Caceres:2019pgf}
appeared on the arXiv. They also suggest a relation between terms in subregion complexity and entanglement
entropy. In particular, in their eqs. (7.8) and (7.9), they guess (supported by numerics) some expressions
for subregion CA and CV 2.0 for a segment in global AdS$_3$.
These expressions should be connected via analytic continuation
to our calculations for  subregion complexity of a segment  in BTZ,  eq. (\ref{CABTZ}) and (\ref{total-btz-bulk}).


\section{Subregion complexity for a segment in AdS$_3$}
\label{sect:AdS}

 It is useful to review the AdS$_{3}$ calculation \cite{Carmi:2016wjl,Chapman:2018bqj,Caceres:2019pgf}
to set up the notation and the procedure, and as a warm-up for the more complicated BTZ case.
because afterwards we will be interested in the more complicated BTZ case.
 Let us consider gravity with negative cosmological constant in $2+1$ dimensions
\beq
S=\frac{1}{16 \pi G} \int \le R+\frac{2}{L^2} \ri   \sqrt{-g} \, d^3 x  \, ,
\label{HIlagrangian}
\eeq
which has as a solution AdS$_3$ spacetime,
whose metric in Poincar\'e coordinates reads
\beq
ds^2 = \frac{L^2}{z^2} \le -dt^2 + dz^2 + dx^2 \ri \, .
\eeq
The AdS curvature is $R=-6/L^2$ and $L$ is the AdS length.
The central charge of the dual conformal field theory is:
\beq
c=\frac{3 L}{2 G} \, .
\eeq

\begin{figure}[h]
\centering
\includegraphics[scale=0.7]{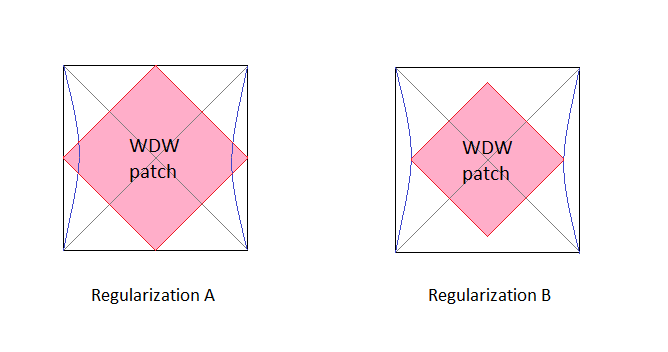}
\caption{The two  regularizations commonly used in the CA conjecture.}
\label{2regs}
\end{figure}

Two common regularizations \cite{Carmi:2016wjl} are used in the CA conjecture 
(see figure \ref{2regs}): 
\begin{itemize}
\item Regularization $A$:
 the WDW patch is built 
starting from the boundary $z=0$ of the spacetime and a cutoff is then introduced
at $z=\varepsilon$.
\item Regularization B: the WDW patch is built from the surface $z=\varepsilon$.
\end{itemize}
We will mostly use regularization $B$;  comparison with regularization $A$
is discussed in Appendix  \ref{app:other-reg}.

\begin{figure}[h]
\center
\begin{tabular}{cc}
\includegraphics[scale=0.5]{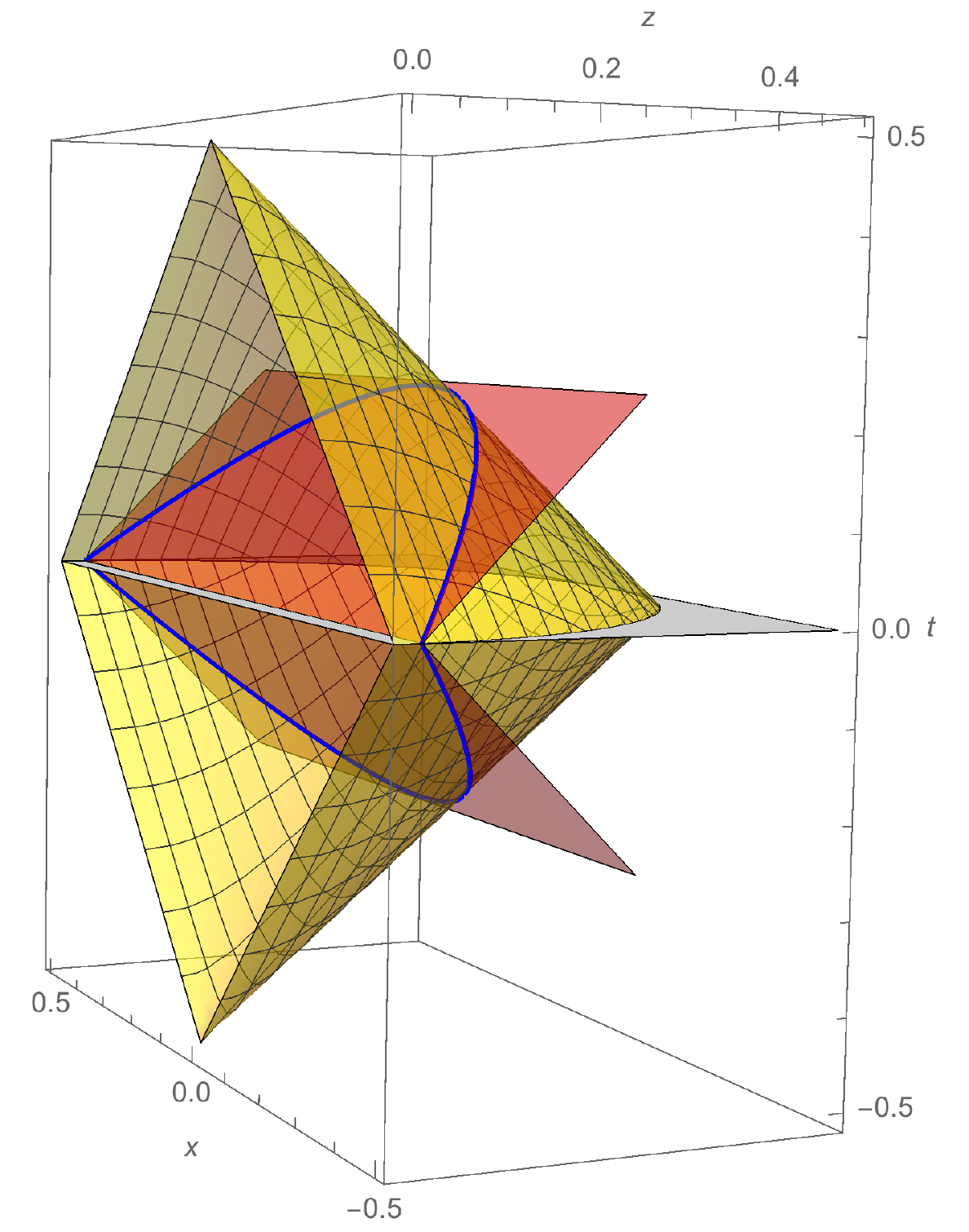} & \includegraphics[scale=0.55]{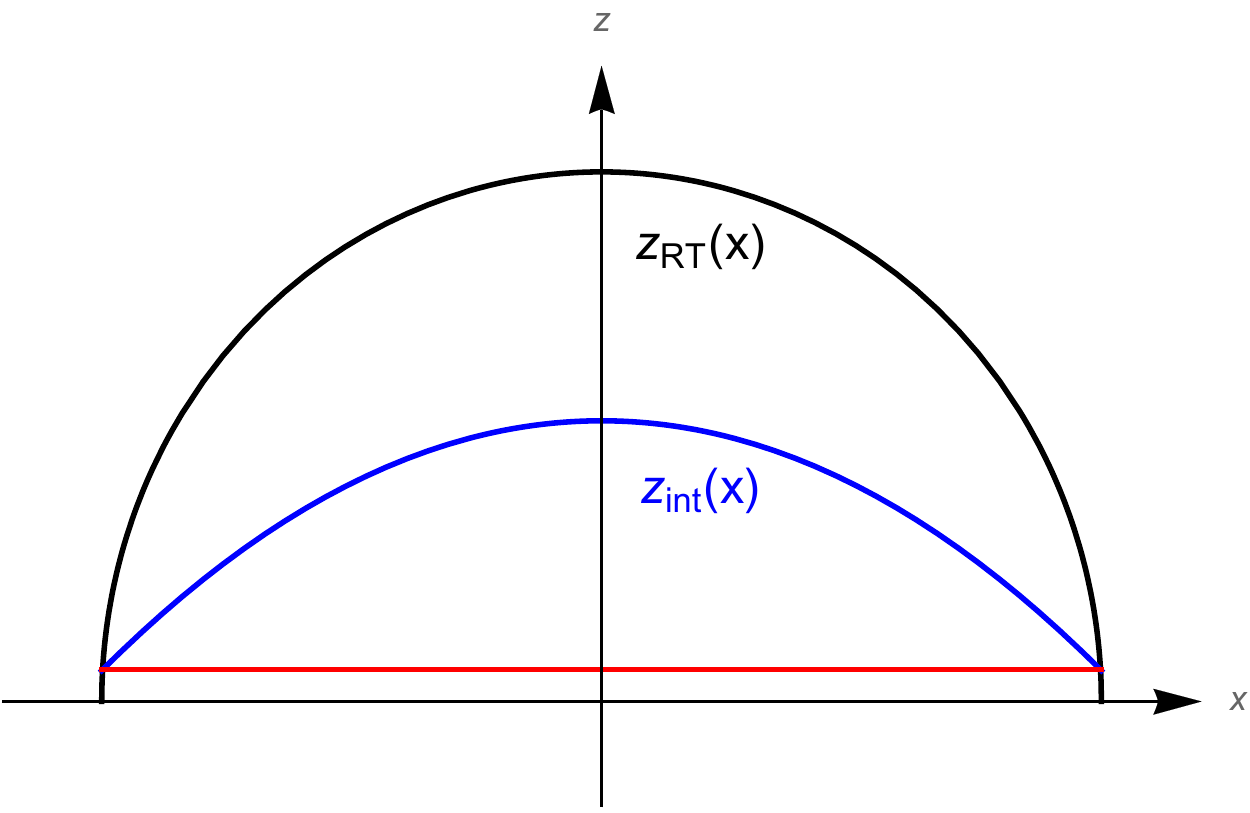}
\end{tabular}
\caption{Left: Intersection of WDW patch  with entanglement wedge in the $(x,z,t)$ space.
The boundary of the entanglement wedge is in yellow,
while the boundary of the WDW patch is in red.
Right: intersections in the $(x,z)$ plane, with $z_{RT}$ in black, $z_{int}$ in blue
and the cutoff $z=\varepsilon$ in red.}
\label{ads}
\end{figure}

We consider a subregion on the boundary given by a strip of length $l$ and
for convenience we take $x \in \left[ - \frac{l}{2}, \frac{l}{2} \right]$,
  at the constant time slice $t=0$.
The geometry relevant to the computation of  action complexity 
is the  intersection between  the entanglement wedge  \cite{Headrick:2014cta}  of the subregion
with the WDW patch \cite{Brown:2015bva,Brown:2015lvg}, see figure \ref{ads}.
We will consider all the contributions to the action 
involving null surface and joint terms introduced in \cite{Lehner:2016vdi}.
The intersection point between the
 WDW patch, the entanglement wedge and the boundary at $z=0$, $x=\pm l/2$
gives a codimension-3 joint,  that a priori can contribute. 
This kind of joint exists just in regularization $B$;
we will check that regularization $A$ gives a similar result in Appendix \ref{app:other-reg}.
So we believe that this joint at most shifts the action of an overall constant.

We use regularization $B$ with a cutoff a $z=\varepsilon$.
The  Ryu-Takayanagi (RT) surface \cite{Ryu:2006bv} is given by the space-like geodesic:
\beq
t=0 \, , \qquad
z^2 + x^2 = \le  \frac{l}{2} \ri^2 \, ,
\eeq
which is a circle of radius $ l/2$.
It is convenient to introduce:
\beq
z_{RT}=\sqrt{\left( \frac{l}{2} \right)^2 -x^2} \, .
\eeq
The entanglement wedge is a cone whose null boundaries are parameterized by
\beq
\label{twedge-ads}
t_{\rm EW} = \pm \le \frac{l}{2} - \sqrt{z^2 + x^2}   \ri \, .
\eeq
The boundaries of the WDW patch,
 which are attached to the regulator surface, are described by the equations
\beq
t_{\rm WDW} = \pm \le z- \varepsilon  \ri \, .
\eeq
The intersection curve between the null boundary  of the WDW patch and the one of the entanglement wedge  is
\beq
z_{\rm int}= \frac{(l + 2 \varepsilon)^2 -4 x^2 }{4 (l +2 \varepsilon) } \, 
\qquad {\rm or} \qquad
x_{\rm int} = \frac{1}{2} \sqrt{(l+2 \varepsilon ) (l-4 z+2 \varepsilon )} \, .
\label{intersez}
\eeq
The  UV cutoff $\varepsilon$ for the radial coordinate $z$
intersects the RT surface at the following value of $x$:
\beq
x_{\rm max} = \sqrt{\le \frac{l}{2} \ri^2 - \varepsilon^2}  \, .
\label{x max AdS}
\eeq
This shift from $x=l/2$ is necessary for a correct regularization
of the on-shell action.
Following \cite{Lehner:2016vdi}, the action includes several terms
\beq
I=I_{\rm bulk}+I_{\rm b}+I_{\rm ct}+ I_{\mathcal{J}} \, ,
\eeq
where $I_{\rm bulk}$ is the bulk term (see eq. \ref{HIlagrangian}), 
$I_{\rm b}$ the null boundary term (see (\ref{null-bou})), 
$I_{\rm ct}$ the counterterm (\ref{counterterm}) and
$I_{\mathcal{J}}$ the null joint contribution (\ref{joint}).

\subsection{Bulk term}

The curvature is constant and so the Einstein-Hilbert term (\ref{HIlagrangian})
is proportional to the spacetime volume.
We can split the bulk contribution in two parts, based on the intersection between 
the WDW patch and the entanglement wedge, which we parametrize with the function $ z_{int} (x) . $
 In the first region the WDW patch is subtended by the entanglement wedge. 
 Consequently, we integrate along time $ 0 \leq t \leq t_{\rm WDW} (z)$, 
  then the radial direction along $ \varepsilon \leq z \leq z_{\rm int} (x)$,  and finally along the coordinate $ 0 \leq x \leq x_{\rm max}$:
 \beq
 I^1_{\rm bulk}  = - \frac{L}{4 \pi G} \int_{0}^{x_{max}} dx \int_{\varepsilon}^{z_{int}} dz \int_{0}^{t_{WDW}} dt \, \frac{1}{z^{3}} 
 \label{integralozzo1}
 \eeq
In the second region the entanglement wedge is under the WDW patch, 
 then the integration involves the endpoints $ 0 \leq t \leq t_{\rm EW}(z,x), z_{\rm int} (x) \leq z \leq z_{\rm RT} (x) $
 and finally $ 0 \leq x \leq x_{\rm max}$ :
  \beq
I^{2}_{\rm bulk}  = - \frac{L}{4 \pi G} 
\int_{0}^{x_{max}} dx \int_{z_{int}}^{z_{RT}} dz \int_{0}^{t_{EW}} dt \, \frac{1}{z^{3}} 
 \label{integralozzo2}
 \eeq
A direct evaluation of the integrals gives: 
   \bea
I^1_{\rm bulk} &=&  - \frac{L}{16 \pi G} \frac{l}{\varepsilon} - \frac{L}{4 \pi G} \log \le  \frac{\varepsilon}{l} \ri
- \frac{L}{8 \pi  G} \, .
\nl
I^{2}_{\rm bulk} &=&  \frac{L}{8 \pi G} \log \le \frac{\varepsilon}{l} \ri
+ \frac{L( \pi ^2 +8)}{64 \pi  G} \, .
\eea
The total result of the bulk action is: 
\beq
 I_{\rm bulk}^{\rm AdS} = 4 (I_{\rm bulk}^1 + I_{\rm bulk}^2) 
 = - \frac{L}{4 \pi G} \frac{l}{\varepsilon} + \frac{L}{2 \pi G} \log \le \frac{l}{\varepsilon} \ri + \frac{L \pi}{16 G}
 \, . \label{total-bulk-ads}
\eeq

\subsection{Null boundary counterterms}

A hypersurface described by the scalar equation $\Phi(x^a)=0$ has a normal vector $k_a= - \p_a \Phi$.
If the hypersurface is null, $k_a k^a=0$ and then it can be shown \cite{Poisson:2009pwt}
 that the hypersurface is generated by null geodesics,
which have $k^\a$ as a tangent vector.

In correspondence of a null boundary,
the following term should be added to the action \cite{Lehner:2016vdi}:
\beq
I_b= \int dS \, d\lambda \,  \sqrt{\sigma}
\kappa \, ,
\label{null-bou}
\eeq
where $\lambda$ is the geodesic parameter, $S$ the transverse spatial directions,
$\sigma$ is the determinant of the induced metric on $S$
and $\kappa$ is defined by the geodesic equation
\beq
k^\mu D_\mu k^\a =\kappa \,  k^\a \, .
\eeq

In our case, the null normals to the WDW patch and the entanglement wedge
 are given respectively by the following 1-forms:
\beq
\mathbf{k}^{\pm}= \alpha \le \pm dt -dz  \ri \, , \qquad
\mathbf{w}^{\pm} = \beta \le \pm dt + \frac{z dz}{\sqrt{z^2 + x^2}} + \frac{x dx}{\sqrt{z^2 + x^2}} \ri \, ,
\eeq
where $\a,\b$ are arbitrary constants that will cancel in the final result.
We denote by $({k}^{\pm})^\mu$ and $({w}^{\pm})^\mu$ the corresponding vectors.
It can be checked that they correspond to an affine
 parametrization of their null surfaces, i.e.
 \beq
(k^\pm)^\mu  D_\mu (k^\pm)^\a =0 \, , \qquad
(w^\pm)^\mu  D_\mu (w^\pm)^\a =0 \, .
\label{geo-geo}
 \eeq
 The term (\ref{null-bou}) vanishes in our calculation because we used
an affine parameterization, see eq. (\ref{geo-geo}).

We still need to include the contribution from the counterterm, 
which ensures the reparameterization invariance of the action:
\begin{equation}
\label{counterterm}
I_{\rm ct} = \frac{1}{8 \pi G} \int d\lambda \, dS \, \sqrt{\sigma} \, \Theta \, \log \left| \tilde{L} \, \Theta \right| \, ,
\end{equation}
where $\Theta$ is the expansion scalar of the boundary geodesics
and $\tilde{L}$ is an arbitrary scale.
If an affine parameterization is used, we can use the result \cite{Poisson:2009pwt}
\begin{equation}
\label{expansion}
\Theta
= D_{\mu} k^{\mu} \, .
\end{equation}

We can then evaluate eq. (\ref{counterterm}) on each boundary:
\begin{itemize}
\item
The counterterm on the entanglement wedge boundary
vanishes because  $\Theta=0$. 
This agrees with the calculations in \cite{Headrick:2014cta}.

\item
For the boundary of the WDW patch we obtain:
\bea
I_{\rm ct}^{\rm WDW}  &= &  - \frac{L}{2 \pi G} \int_0^{x_{\rm max}} dx \int_{\varepsilon}^{z_{\rm int}} \frac{dz}{z^2} \log \left| \alpha
\frac{\tilde{L} z}{L^2} \right|  \nl
& = & \frac{L}{4 \pi G} \frac{l}{\varepsilon} \left[ 1 + \log \le \alpha \frac{\tilde{L}\varepsilon}{L^2} \ri \right]  
+ \frac{L}{4 \pi G} \log \le \frac{\varepsilon}{l} \ri \log \le \alpha^2 \frac{\varepsilon l \tilde{L}^2}{L^4} \ri 
\nl
&+& \frac{L}{2 \pi G} \log \le \frac{\varepsilon}{l} \ri
+ \frac{L \pi}{12 G}  \, .
\label{controt}
\eea

\end{itemize}

\subsection{Joint terms}

The  contribution to the gravitational action 
coming from a codimension-$2$ joint, 
given by intersection of two codimension-$1$ null surfaces \cite{Lehner:2016vdi}, is 
\begin{equation}
\label{joint}
I_{\mathcal{J}} = \frac{\eta}{8 \pi G} \int_{-x_{max}}^{x_{max}} dx \sqrt{\sigma} 
\log \left| \frac{\mathbf{a_{1}} \cdot \mathbf{a_{2}}}{2} \right|
\end{equation} 
where $\sigma$ is the induced metric determinant on the codimension-$2$ surface, 
$\mathbf{a_{1}}$ and $\mathbf{a_{2}}$ are the null normals
 to the two intersecting codimension-$1$ null surfaces and $\eta=\pm 1$.
 The overall sign $\eta$ can be determined as follows:
 if  the  outward direction from a given null surface
 points to the future, we should assign $\eta=1$ if the joint is at the future of the null surface,
 and $\eta=-1$ if it is at the past.
  If the  outward direction from a given null surface
 points to the past,  $\eta=-1$ if the joint is at the future,
 and $\eta=1$ if it is at the past.

The four joints give the following contributions:
\begin{itemize}
\item
The first joint is at the cutoff  $ z= \varepsilon$; we find
\beq
\sqrt{\sigma} = \frac{L}{\varepsilon} 
\, , \qquad
\log \left| \frac{\mathbf{k}^- \cdot \mathbf{k}^+}{2} \right| = \log \left| \alpha^2 \frac{\varepsilon^2}{L^2} \right| \, ,
\eeq
and then  from the general expression (\ref{joint}) 
\beq
I^{\rm cutoff}_{\mathcal{J}} 
= - \frac{L}{4 \pi G} \frac{l}{\varepsilon} \log \le \alpha \frac{\varepsilon}{L} \ri \, .
\eeq

\item
The second joint to compute involves the RT surface:
\beq
\sqrt{\sigma}= \frac{2l L}{l^2 - 4 x^2 } \, , \qquad
\log \left| \frac{\mathbf{w}^+ \cdot \mathbf{w}^-}{2} \right| = \log \left| \beta^2 \frac{l^2 - 4x^2}{4 L^2} \right| \, ,
\eeq
which gives
\beq
I_{\mathcal{J}}^{\rm RT}  
= \frac{L}{4 \pi G} \log \le \frac{\varepsilon}{l} \ri \log \le \frac{\beta^2 \varepsilon l}{L^2} \ri
+ \frac{L \pi}{48 G}  \, .
\eeq

\item
The last joint terms come from the intersections between the 
null boundaries of the WDW patch and the ones of the entanglement wedge:
\beq
\sqrt{\sigma} =  \frac{4L (l+2 \varepsilon)}{(l-2x+2 \varepsilon)(l+2x + 2 \varepsilon)}  \, ,  
\eeq
\beq
\log \left| \frac{\mathbf{k}^+ \cdot \mathbf{w}^+}{2} \right| = \log \left|  \frac{(l-2x+2\varepsilon)(l+2x+2\varepsilon)}{4L(4x^2 + (l+2\varepsilon)^2)} \right|^2 \, .
\eeq
Therefore the joints evaluate to
\beq
I^{\rm int}_{\mathcal{J}} 
 = - \frac{L}{2 \pi G} \log \le \frac{\varepsilon}{l} \ri  \log \le \frac{\alpha \beta}{2} 
 \frac{\varepsilon l}{L^2} \ri - \frac{5 \pi L}{48  G} \, .
\eeq
\end{itemize}

Summing all the joint contributions we find
\beq
I_{\mathcal{J}}^{\rm tot} = - \frac{L}{4 \pi G} \frac{l}{\varepsilon} \log \le \alpha \frac{\varepsilon}{L} \ri 
+ \frac{L}{4 \pi G} \log \le \frac{\varepsilon}{l} \ri \log \le \frac{4 L^2}{\alpha^2 \varepsilon l} \ri
- \frac{\pi L}{12 G}  \, . 
\label{jj-tot}
\eeq
Note that the dependence on the normalization constant 
$\beta$ of the normals cancels in (\ref{jj-tot}); this is due to the fact that the null surfaces 
which have the RT surface as boundaries have vanishing expansion parameter $\Theta$.
Also, when summing the joint term (\ref{jj-tot}) with the counterterm contribution (\ref{controt})
 the double log terms  cancel and the dependence on $\a$ cancels.

\subsection{Complexities}
Summing all the contributions, the action complexity is:
\beq
\mathcal{C}_A^{\rm AdS}=\frac{ I_{\rm tot}^{\rm AdS}}{\pi}=
\frac{c}{3 \pi^2 } \left\{
\frac{l}{2 \varepsilon} \log \le \frac{\tilde{L}}{L} \ri 
- \log \le \frac{2 \tilde{L}}{L} \ri \log \le \frac{l}{\varepsilon} \ri + \frac{\pi^2}{8} \right\} \, .
\label{AdS-CA}
\eeq
Instead, from eq. (\ref{total-bulk-ads}), the spacetime volume complexity is:
\beq
\mathcal{C}_{V 2.0}^{\rm AdS}
 = \frac{2}{3} c \left\{
\frac{l}{\varepsilon} - 2\log \le \frac{l}{\varepsilon} \ri - \frac{ \pi^2}{4 } \right\}
 \, . 
 \label{AdS-CV2}
\eeq
Both the calculations are in agreement with \cite{Caceres:2019pgf}.
In both the expressions for the complexity we recognize
a term proportional to the entanglement entropy of the segment:
\beq
S^{\rm AdS}=\frac{c}{3} \log \left(
\frac{l }{\varepsilon}  \right)  \, .
\eeq
This suggests that the complexity for a single interval 
has a leading divergence  proportional to the length of the subregion on the boundary, 
a subleading divergence proportional to the entanglement entropy and a constant finite piece.
We test this expression for the BTZ case in the next section.


\section{Subregion complexity for a segment in the BTZ black hole}
\label{sect:BTZ}

\begin{figure}[h]
\center
\begin{tabular}{cc}
\includegraphics[scale=0.5]{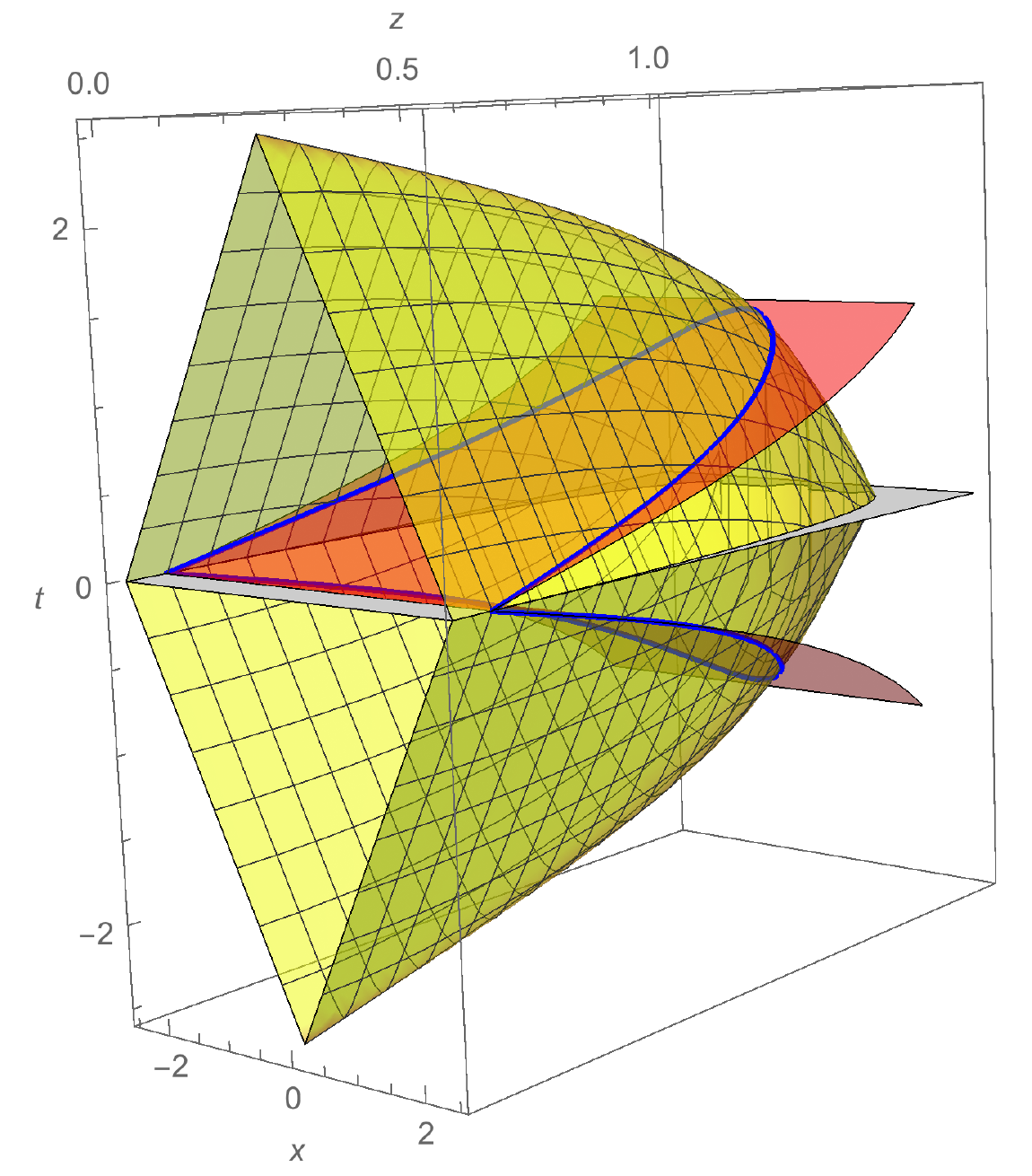} & \includegraphics[scale=0.55]{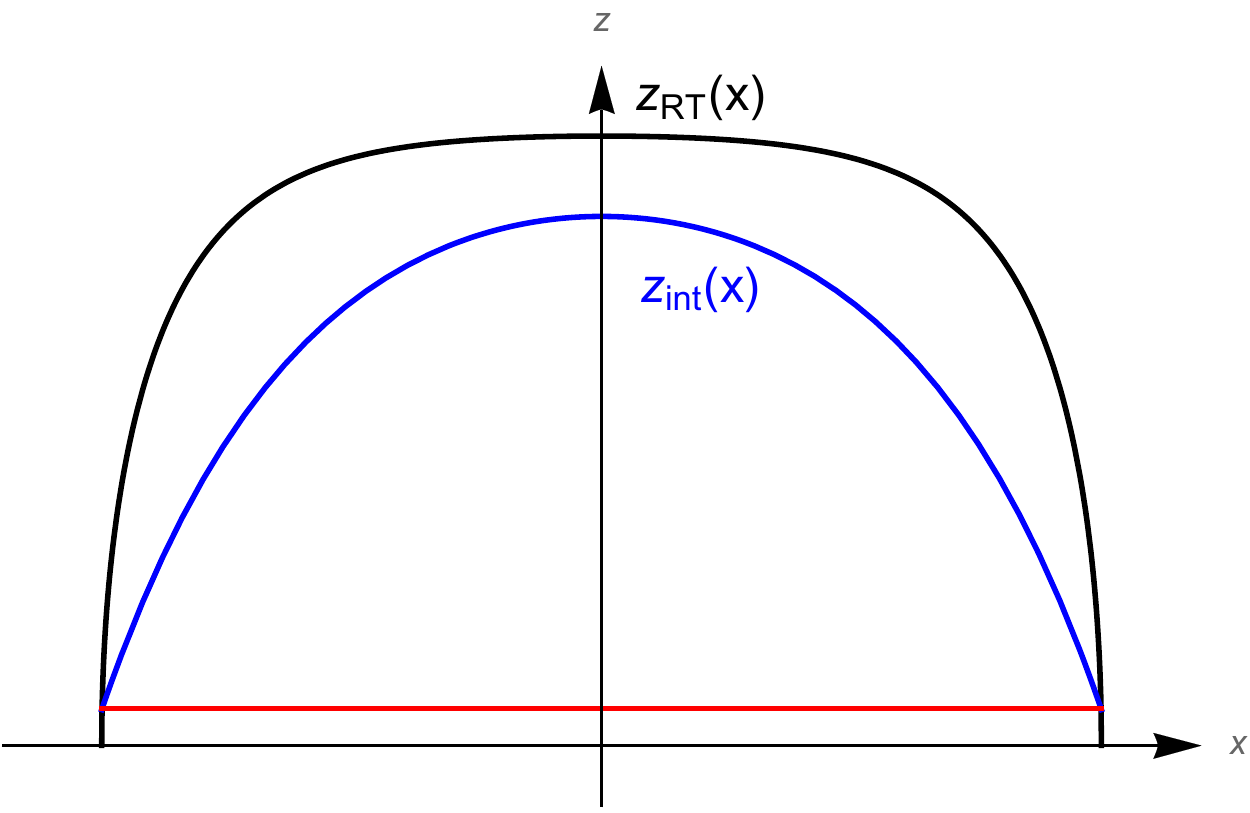}
\end{tabular}
\caption{ Region relevant to the action computation for a segment in the BTZ case, for $l=5$.
Left: Intersection of WDW patch  with entanglement wedge in the $(x,z,t)$ space.
The boundary of the entanglement wedge is in yellow,
while the boundary of the WDW patch is in red.
Right: intersections in the $(x,z)$ plane, with $z_{RT}$ in black, $z_{int}$ in blue
and the cutoff $z=\varepsilon$ in red.}
\label{fig-BTZ}
\end{figure}

We consider the metric of the planar BTZ black hole in $2+1$ dimensions with non-compact coordinates $(t,z,x)$ 
\beq
\label{metric}
ds^2 = \frac{L^2}{z^2} \le - f dt^2 + \frac{dz^2}{f} + dx^2 \ri \, ,
\qquad f = 1- \le \frac{z}{z_h} \ri^2 \, ,
\eeq
where $L$ is the $\mathrm{AdS}$ radius and $z_h$ is the position of the horizon.
The mass, the temperature and the entropy are:
\beq
M = \frac{L^2}{8 G z_h^2} \, , \qquad
T = \frac{1}{2 \pi z_h} \, , \qquad
S= \frac{\pi L^2}{2 G z_h} \, .
\eeq
The geometry needed to evaluate the subregion complexity for a segment is shown in figure
\ref{fig-BTZ}

The RT surface
 is a spacelike geodesic which lies on a constant time slice $t=0$ and which is
  anchored at the edges of the boundary subregion \cite{Balasubramanian:2011ur}:
\begin{equation}
\label{RT surface}
x_{\pm}(z)= \frac{1}{4} z_h \left[ \log \le \frac{J+1}{J-1} \ri^2 
+  \log \le \frac{z_h^2 -J z^2 \pm \sqrt{z_h^4 - \left( 1+J^{2}  \right) z_h^2 z^{2}+J^{2} z^{4}}}
 {z_h^2 + J z^2 \pm \sqrt{z_h^4 - \left( 1+J^{2}  \right) z_h^2 z^{2}+J^{2} z^{4}}} \ri^2 \right] \, ,
\end{equation}
where
\begin{equation}
J = \coth \left( \frac{l}{2 z_h} \right) \, .
\end{equation}
The turning point of the geodesic is at $x_{\pm} (z_{*})=0$, where
\begin{equation}
z_{*}= z_h \tanh \left( \frac{l}{2 z_h} \right) \, .
\end{equation}
Since $z_{*} < z_h $ for every value of the boundary subregion size $l$,
 the geodesic never penetrates inside the event horizon of the black hole. 
It is convenient to invert eq. (\ref{RT surface}):
\begin{equation}
\label{zgeod}
 z_{RT} =
 z_h \sqrt{\frac{\cosh \left( \frac{l}{z_h} \right) - 
 \cosh \left( \frac{2x}{z_h} \right)}{\cosh \left( \frac{l}{z_h} \right) +1}} \, .
\end{equation}

In our static case, the entanglement wedge coincides with the causal wedge  \cite{Czech:2012bh,Hubeny:2012wa,Wall:2012uf}, 
which can be  constructed by sending null geodesics from
 the causal diamond on the boundary into the bulk. 
 The explicit expressions of such geodesics 
 are \cite{Hubeny:2012wa}
\bea
 \tilde{x}_{\rm EW} (z,j) &=&  \frac{z_h}{2} \log  \le \frac{\sqrt{z_h^2 + j^2 (z^2 - z_h^2)}+ j z}{\sqrt{z_h^2 + j^2 (z^2 - z_h^2)}- j z} \ri \, ,  \nl
 \tilde{t}_{\rm EW} (z,j)&=& \pm \left[ \frac{l}{2} + \frac{z_h}{2} \log   \le \frac{\sqrt{z_h^2 + j^2 (z^2 - z_h^2)}- z}{\sqrt{z_h^2 + j^2 (z^2 - z_h^2)}+z} \ri \right] \, .
\label{t,x wedge}
\eea
We obtain an analytical expression for the boundary of the entanglement 
wedge in terms of a unique explicit relation between $(t,z,x)$ by determining $j=j(z,x)$ 
from the first equation in  (\ref{t,x wedge})
and then inserting it 
 into the second equation of
   (\ref{t,x wedge}). The result
   can be written as
\begin{equation}
\label{twedge}
t_{\rm EW} =\pm \left[
 \frac{l}{2} - z_h \, \mathrm{arccoth}  \le \frac{\sqrt{2} z_h \cosh \le \frac{x}{z_h} \ri}{\sqrt{2z^2 + z_h^2 \cosh \le \frac{2x}{z_h} \ri - z_h^2}}  \ri \right]
 \, .
\end{equation}

The WDW patch is delimited by the radial null geodesics:
\beq
t_{\mathrm{WDW}} =\pm \frac{z_h}{4} 
\log \le \frac{z_h+z}{z_h-z} \frac{z_h-\varepsilon}{z_h+\varepsilon}
\ri^2 \, .
\eeq
The intersection between the boundary of the WDW patch and the entanglement wedge is:
\beq
\label{xintpos}
t_{\rm int}=t_{\mathrm{WDW}} \, ,  \qquad
 z_{\rm int} = z_h \frac{\cosh \left[ \frac{l}{2 z_h} + \mathrm{arctanh} \le \frac{\varepsilon}{z_h} \ri  \right] - \cosh \le \frac{x}{z_h} \ri}{\sinh \left[ \frac{l}{2 z_h} + \mathrm{arctanh} \le \frac{\varepsilon}{z_h} \ri   \right]} \, . 
 \eeq
We plot this curve in Fig. \ref{fig-BTZ}. 

As in the AdS case, we denote by $ x_{\rm max} $ the maximum value of the transverse coordinate, 
which is reached when we evaluate the RT surface at $ z= \varepsilon$:
\beq
x_{\rm max} = 
z_h \, \mathrm{arccosh}  \left[ \sqrt{1- \frac{\varepsilon^2}{z_h^2}} \, \cosh \le \frac{l}{2 z_h} \ri \right] \, .
\label{xmax}
\eeq

\subsection{Bulk contribution}
\label{sect-Bulk contribution 2nd regularization}

We split the integration region as in the AdS case, see
eqs. (\ref{integralozzo1},\ref{integralozzo2}).
The total bulk action then is $I_{\rm bulk} = 4 ( I^1_{\rm bulk}  + I^2_{\rm bulk}) $. 
A direct calculation gives:
\beq
\begin{aligned}
I_{\rm bulk}   =  \frac{L}{8 \pi G z_h} & \int_{0}^{x_{max}(\varepsilon)} dx \, \left\lbrace \frac{4 \sinh \left[ \frac{l}{2 z_h} + \mathrm{arctanh}  \le \frac{\varepsilon}{z_h} \ri  \right]}{\cosh \le \frac{l}{2 z_h} + \mathrm{arctanh} \le \frac{\varepsilon}{z_h} \ri \ri - \cosh \le \frac{x}{z_h} \ri }  -\frac{4 z_h}{ \varepsilon }  \right. \\
& \left.  + 2 \coth \le \frac{x}{z_h} \ri \log \left| \frac{\sinh \le \frac{l-2x}{2 z_h} \ri  \sinh^2 \left[ \frac{l+2x+2z_h \, \mathrm{arctanh} (\varepsilon/z_h)}{4 z_h} \right]}{\sinh \le \frac{l+2x}{2 z_h} \ri  \sinh^2 \left[ \frac{l-2x+2z_h \, \mathrm{arctanh} (\varepsilon/z_h)}{4 z_h} \right]}  \right| \right\rbrace \, .
\end{aligned}
\label{final integrand bulk}
\eeq
This integral can be computed analytically, and gives the CV 2.0
complexity in eq (\ref{total-btz-bulk}).

\subsection{Null normals}
In order to compute the counterterms due to the null surfaces and
the joint contributions, the null normals are needed.
It is convenient to use an affine parameterization, which can
be found using the following Lagrangian description of geodesics:
\begin{equation}
\label{lagrangian}
\mathcal{L} = \frac{L^2}{z^2} \le - f(z) \, \dot{t}^{2} + \frac{\dot{z}^{2}}{f(z)} + \dot{x}^{2} \ri
\end{equation}
where the dot represents the derivative with respect to the affine parameter $\lambda$. 
Since the Lagrangian does not depend on $t$ and $x$, we have two constants of motion
\begin{equation}
\label{motion}
E = - \frac12 \frac{\p \mathcal{L}}{\p \dot{t}} = \frac{L^2}{z^2} f(z) \, \dot{t} \, , \qquad J = \frac12 \frac{\p \mathcal{L}}{\p \dot{x}} =\frac{L^2}{z^{2}} \dot{x} \, .
\end{equation}
Imposing the null condition $\mathcal{L} = 0$ and making use of eq. (\ref{motion}) leads to
\begin{equation}
\label{zdot}
\dot{z} = \pm \, \frac{z^2}{L^2} \, \sqrt{E^{2} - J^{2} f(z)} \, .
\end{equation}
Therefore, from eqs. (\ref{motion}) and (\ref{zdot}), the tangent vector to the null geodesic is
\begin{equation}
\label{nullvector}
V^{\mu} = \left( \dot{t}, \, \dot{z}, \, \dot{x} \right) = \left(\frac{z^2}{L^2 f(z)} E , \, \pm \, \frac{z^2}{L^2} \, \sqrt{E^{2} - J^{2} f(z)} , \, \frac{z^2}{L^2} \, J \right)  \, .
\end{equation}
Lowering the contravariant index with the metric tensor, we get the normal $1$-form to the null geodesic
\begin{equation}
\label{nullform}
\mathbf{V} = V_{\mu} dx^{\mu} = - E \, dt \, \pm \, \frac{\sqrt{E^{2} - J^{2} f(z)}}{f(z)} \, dz + J \, dx \, .
\end{equation}

The null geodesics which bound the WDW patch are $x$-constant curves,
and so $J=0$. This gives the following normals:
\begin{equation}
\label{wdwnormals}
\mathbf{k^{+}}= k_{\mu}^{+} dx^{\mu} = \alpha \left( dt - \frac{dz}{f(z)} \right) \, ,
 \qquad \mathbf{k^{-}}= k_{\mu}^{-} dx^{\mu} = \alpha \left( - \, dt - \frac{dz}{f(z)} \right) \, ,
\end{equation}
where $\a$ is an arbitrary constant.

The null geodesics that bound the entanglement wedge are normal to the RT surface,
i.e.
\begin{equation}
V_{\mu} \, \frac{dX^{\mu}_{RT} (x)}{dx} = 0 \, , \qquad 
X^{\mu}_{\rm RT} (x) = \left( 0, z_{RT}, \, x \right) \, ,
\label{rtvector}
\end{equation}
where $z_{RT}$ is given in eq. (\ref{zgeod}). 
With this condition and eqs. (\ref{nullform}) and (\ref{rtvector}), 
we find a relation between the two constants of motion $E$ and $J$ which then gives (for $t>0$ and $t<0$ respectively)
\beq
\mathbf{w^{\pm}} = w_{\mu}^{\pm} dx^{\mu} =
\beta \left( \pm dt + a \, dz +b \, dx \right)  \, , 
\eeq
where
\beq
a = \frac{e^{-\frac{x}{z_h}} \left( e^{\frac{2x}{z_h}}+1 \right) z z_h^2}{\left(z_h^2-z^{2} \right) \sqrt{4 z^2 + e^{- \frac{2x}{z_h}} \le e^{\frac{2x}{z_h}} -1 \ri^2 z_h^2}} \, , \qquad
b=\frac{e^{-\frac{x}{z_h}} \left( e^{\frac{2x}{z_h}}-1 \right) z_h}{\sqrt{4 z^2+ e^{- \frac{2x}{z_h}} \le e^{\frac{2x}{z_h}} -1 \ri^2 z_h^2}} \, .
\eeq

\subsection{Null boundaries and counterterms}
\label{sect-BTZ-countertems}

The term in eq. (\ref{null-bou}) vanishes because we used an affine parameterization.
The counterterm in eq. (\ref{counterterm}) gives:
\begin{itemize}
\item For the null normals of the boundary of the entanglement wedge, 
this contribution vanishes because $\Theta=D_{\mu} (w^{\pm})^{\mu} = 0$.
\item For the null normals of the boundary of the WDW patch, a direct calculation gives $\Theta = \frac{\alpha z}{L^2}$ and:
\beq
\begin{aligned}
I_{\rm ct}^{\rm WDW} & = - \frac{L}{2 \pi G} \int_0^{x_{\rm max}} dx \int_{\varepsilon}^{z_{\rm int} (x)} dz \, \frac{1}{z^2} \log \left| \frac{\tilde{L}}{L^2} \, \alpha z \right| = \\
& =  \frac{L}{2 \pi G} \int_0^{x_{\rm max}} dx  \, \left\lbrace \frac{1 + \log \left| \frac{\tilde{L}}{L^2} \, \alpha \varepsilon \right|}{\varepsilon} + \frac{\sinh \le \frac{l}{2 z_h} + \mathrm{arctanh} \le \frac{\varepsilon}{z_h} \ri \ri}{z_h \left[ \cosh \le \frac{x}{z_h} \ri - \cosh \le \frac{l}{2 z_h} \ri + \mathrm{arctanh} \le \frac{\varepsilon}{z_h} \ri  \right]}  \times \right. \\
& \left. \times \le 1 + \log \left| \frac{\tilde{L} z_h \alpha}{L^2} \frac{\cosh \le \frac{l}{2 z_h} + \mathrm{arctanh} \le \frac{\varepsilon}{z_h} \ri \ri - \cosh \le \frac{x}{z_h} \ri}{\cosh \le \frac{l}{2 z_h} + \mathrm{arctanh} \le \frac{\varepsilon}{z_h} \ri \ri} \right|  \ri  \right\rbrace \, .
\end{aligned}
\eeq
\end{itemize}

\subsection{Joint contributions}
\label{sect-BTZ-joints}

We evaluate the joint terms in eq (\ref{joint}):
\begin{itemize}
\item The joint at the cutoff gives:
\begin{equation}
I_{\mathcal{J}}^{\rm cutoff} = - \, \frac{L}{4 \pi G} \, \int_{0}^{x_{\rm max} } \frac{dx}{\varepsilon} \left| \frac{\alpha^{2} \, z_h^2 \, \varepsilon^{2}}{L^2 (z_h^2-\varepsilon^{2})} \right| \, .
\end{equation}
\item The joint at the RT surface:
\begin{equation}
I_{\mathcal{J}}^{\rm RT} = - \, \frac{L}{4 \pi G z_h} \, \int_{0}^{x_{\rm max} } dx \, \frac{\sinh \le \frac{l}{z_h} \ri}{\cosh \le \frac{l}{z_h} \ri - \cosh \le \frac{2x}{z_h} \ri} \log \left| \frac{\beta^2 z_h^2}{2 L^2} \frac{\cosh \le \frac{l}{z_h} \ri - \cosh \le \frac{2x}{z_h} \ri}{\cosh^2 \le \frac{x}{z_h} \ri} \right| \, .
\label{joint-BTZ-RT}
\end{equation}
\item The joints coming from the intersection between the null boundaries of the 
WDW patch and the ones of the entanglement wedge give:
\begin{equation}
\begin{aligned}
I_{\mathcal{J}}^{\rm int} & =  \frac{L}{2 \pi G z_h} \int_{0}^{x_{\rm max}} dx \, 
\frac{\sinh \le \frac{l}{2 z_h} + \mathrm{arctanh} \le \frac{\varepsilon}{z_h} \ri \ri}{\cosh \le \frac{l}{2 z_h} + \mathrm{arctanh} \le \frac{\varepsilon}{z_h} \ri \ri - \cosh \le \frac{x}{z_h} \ri} \times \\
& \times \log \left| \frac{e^{x/z_h} \alpha \beta z_h^2}{L^2} \frac{\left[ \cosh \le \frac{l}{2 z_h} + \mathrm{arctanh} \le \frac{\varepsilon}{z_h} \ri \ri - \cosh \le \frac{x}{z_h} \ri \right]^2}{ 1 + e^{2x/z_h} \cosh \le \frac{l}{2 z_h} + \mathrm{arctanh} \le \frac{\varepsilon}{z_h} \ri \ri - 2 e^{x/z_h} } \right| \, .
\end{aligned}
\label{joint-BTZ-EW}
\end{equation}
\end{itemize}
All the joints contributions and the counterterm are regularized by the cutoff $\varepsilon$.

\subsection{Complexities}

We performed all the integrals analytically and we
 further simplified the result using various dilogarithm identities, including the relation:
\bea
  8 \, \mathrm{Re} \left[  \mathrm{Li}_2 \le \frac{1 + i e^{\frac{y}{2}}}{1+ e^{\frac{y}{2 }}} \ri 
 - \mathrm{Li}_2 \le \frac{1}{1 + e^{\frac{y}{2 }}} \ri - \mathrm{Li}_2 \le 1 + i e^{\frac{y}{2 }} \ri
   - \mathrm{Li}_2 \le \frac{e^{\frac{y}{2 }} -i}{1+ e^{\frac{y}{2 }}} \ri
 \right] =
\nl
 =  -  \frac{7 \pi^2 }{6 }
+  4 \le \log \le 1 + e^{\frac{y}{2}} \ri \ri^2 
+ \log 2 \left[  2 y - 4  \log \le \frac{e^{y} -1}{y}  \ri  
 + 4 \log \left( \frac{2 }{y} \sinh \frac{y}{2 }  \right) \right] \, ,
\eea
which can be proved by taking a derivative of both side of
the equation with respect to $y$.

The action subregion complexity then is:
\beq
 \mathcal{C}_{A}^{\rm BTZ} = \frac{c}{3 \pi^2 } 
 \left\{ \frac{l}{2 \varepsilon} \log \le \frac{\tilde{L}}{L} \ri -
 \log \le \frac{2 \tilde{L}}{L} \ri
  \log \left( \frac{2 z_h}{\varepsilon} \sinh \le \frac{l}{2 z_h} \ri \right) 
 +\frac{\pi^2}{8} \right\} \, .
\label{total action BTZ simplified}
\eeq
Introducing the entanglement entropy
of an interval
\beq
S^{\rm BTZ}=\frac{c}{3} \log \left(
\frac{2 z_h }{\varepsilon} \sinh \left( \frac{l}{2 z_h} \right)
\right)  \, ,
\label{entanglement entropy BTZ}
\eeq
we can then write it in this form
\beq
 \mathcal{C}_{A}^{\rm BTZ} = \frac{l}{\varepsilon} \frac{c}{6 \pi^2} \log \left(\frac{\tilde{L}}{L} \right)
- \log  \left(\frac{2\tilde{L}}{L} \right) \frac{S^{\rm BTZ}}{\pi^2} + \frac{1}{24} c \, .
\label{ACTION-C}
\eeq
By integration of (\ref{final integrand bulk}), the subregion spacetime complexity is
\beq
\mathcal{C}_{V 2.0}^{\rm BTZ}= \frac{2 c}{ 3 }   \frac{l}{\varepsilon} 
-4 S^{\rm BTZ} - \frac{\pi^2}{6} c \, .
\label{total-btz-bulk}
\eeq
The divergencies of eqs. (\ref{ACTION-C}) and (\ref{total-btz-bulk}) are the same 
as in the AdS case eqs. (\ref{AdS-CA}) and (\ref{AdS-CV2}), which is recovered for
$z_h=0$.

A useful cross-check can be done in the 
 $l  \gg z_h$ limit. Keeping just the terms linear in $l$
 in eq. (\ref{total action BTZ simplified}), we find agreement
 with the subregion complexity $\mathcal{C}_A^{\rm{BTZ},R}$ computed for one side of the Kruskal diagram,
 see \cite{Agon:2018zso,Alishahiha:2018lfv}:
 \beq
\mathcal{C}_A^{\rm{BTZ},R}= \frac{ c}{6 }
\frac{l  }{ \pi^2 } \left[
 \frac{1}{\varepsilon} \log \le \frac{\tilde{L}}{L} \ri
 - \frac{1}{ z_h}  \log \le \frac{2 \tilde{L}}{L} \ri \right] \, .
\label{external total action BTZ}
\eeq
Note that in this limit the $\log \varepsilon$ divergence disappears because
it is suppressed by the segment length $l$.

For comparison, the volume complexity  of an interval for the BTZ
\cite{Alishahiha:2015rta,Abt:2017pmf} is:
\beq
\mathcal{C}_V^{\rm BTZ} =\frac{2 \, c}{3} \le  \frac{l}{\varepsilon} - \pi \, \ri ,
\eeq
and it is non-trivially independent on temperature.
Subregion $\mathcal{C}_V$ at equilibrium is a topologically protected quantity:
for multiple intervals, the authors of \cite{Abt:2017pmf} found the following result
using the the Gauss-Bonnet theorem
\beq
\mathcal{C}_V^{\rm AdS}=  \mathcal{C}_V^{\rm BTZ} = \frac{2 \, c}{3} \le \frac{l_{tot}}{\varepsilon} + \kappa \ri \, ,
\label{CCVA}
\eeq
where $l_{tot}$ is the total length of all the segments and $\kappa$
is the finite part, that depends on topology
\beq
\kappa=- 2 \pi \chi + \frac{\pi}{2} \, m \, ,
\label{CCVB}
\eeq
where $\chi$ is the Euler characteristic  of the extremal surface
(which is equal to $1$ for a disk)
and $m$ is the number of ninety degrees junctions between RT 
surface and boundary segments. 
It would be interesting to see if 
 a similar result could be established for the  CA and CV 2.0 conjecture.
This motivates us to study the two segment case in the next section.



\section{Subregion complexity for two segments in AdS$_3$}
\label{sect:AdS-2seg}

\begin{figure}
\begin{center}
\includegraphics[scale=0.6]{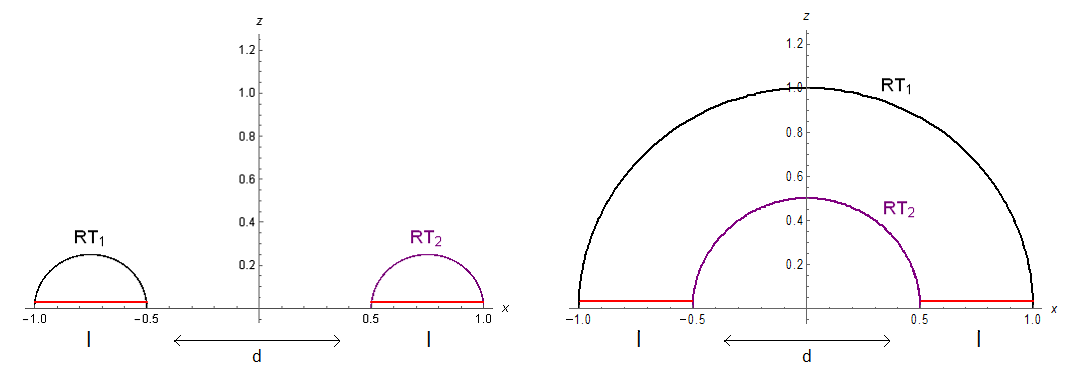}
\caption{The possible RT surfaces for disjoint subregions of length $l=0.5$ with a separation $d=1$, 
on the slice $t=0$. }
\label{RT-disjoint}
\end{center}
\end{figure}

In this section we  evaluate the holographic subregion action complexity 
for a  disjoint subregion on the $AdS_{3}$ spacetime's boundary. 
We consider two segments of size $l$ with a separation equal to $d$,
 located at the spacetime's boundary on the constant time slice $t=0$. 
 For simplicity, we work with a symmetric configuration, 
 in which the two boundary subregions are respectively given by 
 $x \in \left[  -l - d/2 , -d/2 \right]$ and $x \in \left[ d/2 ,l + d/2 \right]$.  
According to the values of the subregions size $l$ and of the separation $d$,
 there are two possible extremal surfaces anchored at the boundary
  at the edges of the two subregions \cite{{Rangamani:2016dms,Headrick:2019eth}}:
\begin{itemize}
 \item The extremal surface (which in this number of dimension is a geodesic)
 is given by the union of the RT surfaces for the individual subregions.
 This is the minimal surface for $d>d_0$, where $d_0$ is a critical distance.
 \item The extremal surface connects the two subregions. 
 This configuration is minimal for $d< d_0$.
\end{itemize} 
The two cases are shown in Fig. \ref{RT-disjoint}. 
The geodesic with the minimal area provides the holographic entanglement entropy 
for the union of the disjoint subregions. 
The critical distance corresponds to the distance for which both the extremal surfaces
have the same  length, i.e.
\beq
d_0=(\sqrt{2}-1) l \, .
\eeq

In the first configuration (see left in Fig. \ref{RT-disjoint}), 
we have two non-intersecting entanglement wedges
and so
\begin{equation}
\label{disjoint-complexity-case1}
\mathcal{C}_A^1=2 \,  \mathcal{C}_A^{\rm AdS} \, , \qquad
\mathcal{C}_{V 2.0}^1=2 \,  \mathcal{C}_{V 2.0}^{\rm AdS} \, .
\end{equation}

\begin{figure}[h]
\center
\begin{tabular}{ccc}
\includegraphics[scale=0.5]{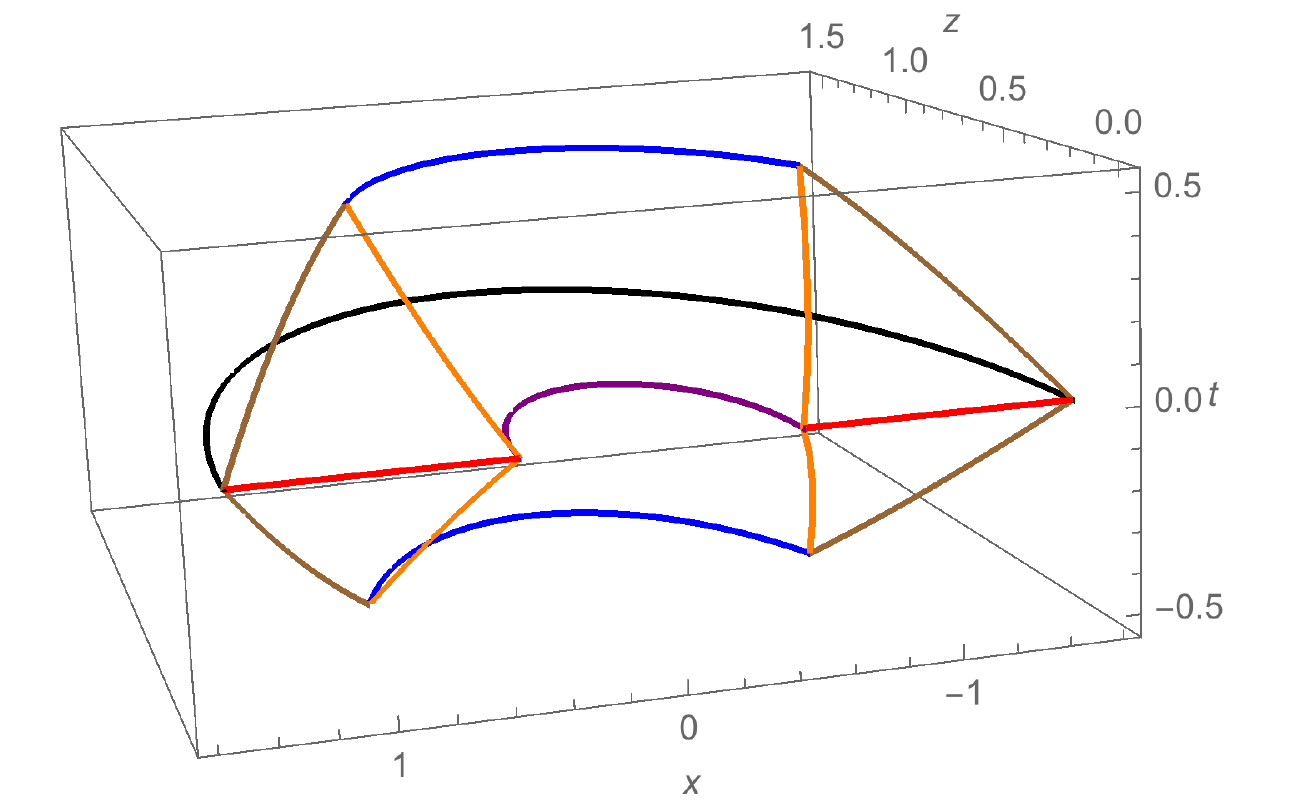} & \includegraphics[scale=0.5]{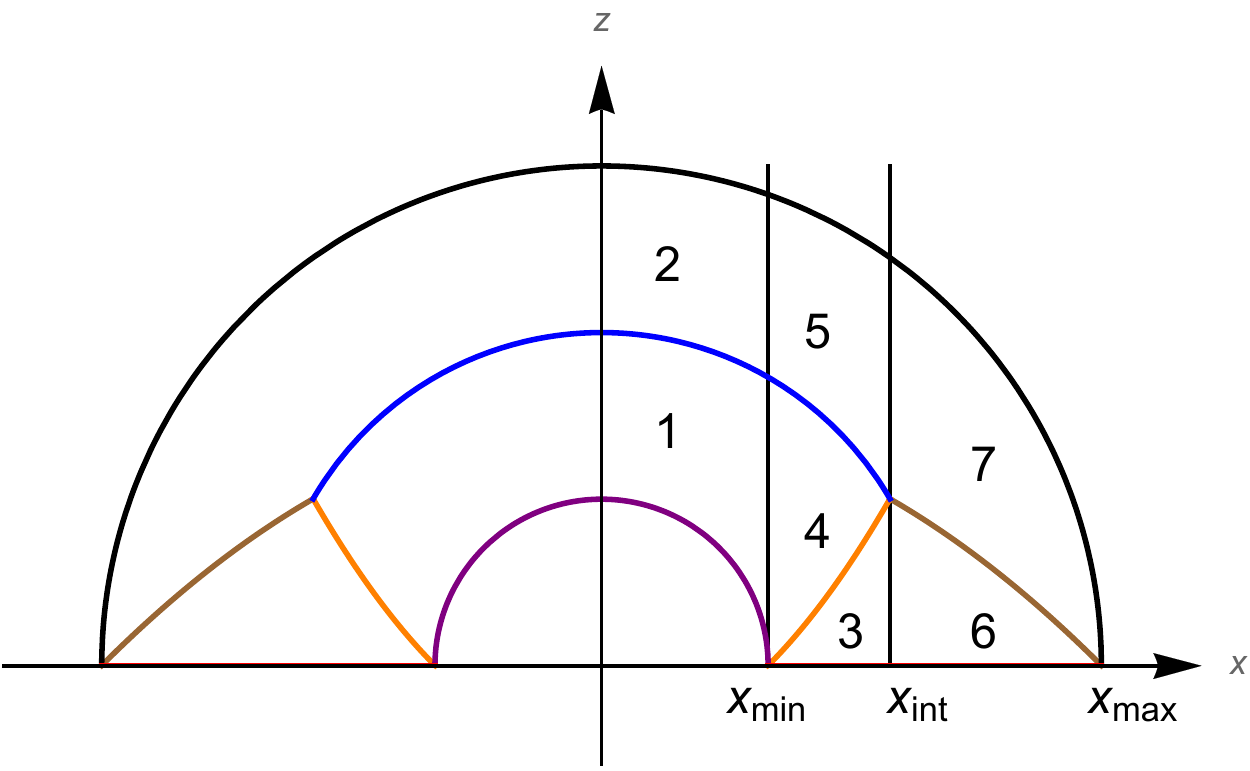} &
 \includegraphics[scale=0.5]{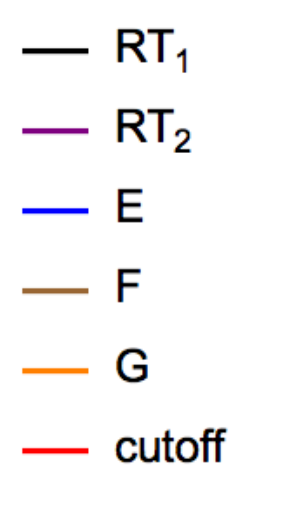}
\end{tabular}
\caption{Left: Bulk region relevant to the action subregion calculation for 
two segments in AdS.
 Right: projection in the $(x,z)$ plane. The regions in which the bulk integral 
 is splitted are numbered. }
\label{geometry-disjoint}
\end{figure}

 For the second configuration (right in Fig. \ref{RT-disjoint}),
 we must perform a new computation. The spacetime region of interest is 
 symmetric both with respect to the $x=0$ slice and to the $t=0$ one. 
As a consequence, we can evaluate the action on the region with $t>0$ and $x>0$
 and introduce opportune symmetry factors. A schematic representation is shown in figure
 \ref{geometry-disjoint}.

The RT surface is the union of the spacelike geodesics anchored 
at the edges of the region $x \in \left[ -l-d/2 , l+d/2 \right]$ and
 $x \in \left[ -d/2 , d/2 \right]$.
 We will denote such geodesics as RT$_1$ and RT$_2$ respectively:
\begin{equation}
\label{zRT-small-big}
z_{RT_1} (x) = \sqrt{\left( \frac{2l+d}{2} \right)^{2} -x^{2}} \, , 
\qquad \qquad  z_{RT_2} (x) = \sqrt{\left( \frac{d}{2} \right)^{2} -x^{2}} \, . 
\end{equation}
With the introduction of the cutoff surface at $z= \varepsilon$, 
RT$_2$ is truncated at $x=x_{min} $ 
and RT$_1$ at $x= x_{max}$, defined by
\beq
x_{min} = \sqrt{\le \frac{d}{2} \ri^2 - \varepsilon^2} \, ,
\qquad
x_{max}= \sqrt{\le \frac{d+ 2 l}{2} \ri^2 - \varepsilon^2} \, .
\eeq
 The null boundaries of the entanglement wedge can be built by 
 sending null geodesics from  RT$_1$ and RT$_2$:
\begin{equation}
\label{tEW}
t_{EW_1}  = \frac{2l+d}{2} - \sqrt{z^{2} + x^{2}} \, ,
\qquad
t_{EW_2} = - \frac{d}{2} + \sqrt{z^{2} + x^{2}} \, .
\end{equation}
The WDW patch, anchored at the cutoff in the present regularization, is bounded by the null surface
\begin{equation}
\label{wdw-ads}
t_{WDW} = z - \varepsilon \, .
\end{equation}

The intersection curve $E$ between the null boundaries of the entanglement wedge,
(built from  RT$_1$ and RT$_2$, see eq. (\ref{tEW})) is 
\begin{equation}
\label{zEW-EW}
t_{E} = \frac{l}{2} \, , \qquad \qquad z_{E} = \frac{1}{2} \sqrt{\left( d+l \right)^{2} -4 x^{2}} \, .
\end{equation}
The intersection $F$ between the boundary of the
WDW patch eq. (\ref{wdw-ads})
and the null surface anchored at RT$_1$ is:
\beq
\label{zWDW-EWbig}
t_{F} = \frac{1}{4} \left[ d + 2 \left( l - \varepsilon \right) - \frac{4 x^{2}}{d+ 2 \left( l + \varepsilon \right)} \right] \, , 
\qquad
 z_{F}  =t_{F}  + \varepsilon \, .
\eeq 
The  intersection $G$
 between the WDW patch eq. (\ref{wdw-ads})
  and the null surface anchored at RT$_2$ gives
\begin{equation}
\label{zWDW-EWsmall}
t_{G}  = - \frac{d}{4} + \frac{x^{2}}{d - 2 \varepsilon} - \frac{\varepsilon}{2} \, , 
\qquad z_{G}  = t_{G}  + \varepsilon\, .
\end{equation}
 The intersection among the three curves described above 
 (obtained solving the condition $z_{E}= z_{F}= z_{G}$) gives
\beq
x_{int} (\varepsilon) = \frac{\sqrt{\le d-2 \varepsilon \ri \left[ d+2 \le l+ \varepsilon \ri \right]}}{2} \, .
\eeq

\subsection{Bulk contribution}

As shown in Fig. \ref{geometry-disjoint}, the total bulk contribution can be divided into 7 parts for computational reasons:
\beq
I_{bulk}= 4 \sum_{i=1}^{7} I_{bulk}^{i} \, ,
\eeq
where
\bea
I_{bulk}^{1} &=& - \frac{L}{4 \pi G} \int_{0}^{x_{min} } dx \int_{z_{RT_2} }^{z_{E} } dz \int_{0}^{t_{EW_2} } \frac{dt}{z^3} \nl
I_{bulk}^{2}  &=&- \frac{L}{4 \pi G} \int_{0}^{x_{min} } dx \int_{z_{E} }^{z_{RT_1} } dz \int_{0}^{t_{EW_1} } \frac{dt}{z^3}  \nl
I_{bulk}^{3} &=& - \frac{L}{4 \pi G} \int_{x_{min} }^{x_{int} } dx \int_{\varepsilon}^{z_{G} } dz \int_{0}^{t_{WDW}} \frac{dt}{z^3} \nl
I_{bulk}^{4} & = &- \frac{L}{4 \pi G} \int_{x_{min} }^{x_{int} } dx  \int_{z_{G} }^{z_{E} } dz \int_{0}^{t_{EW_2} } \frac{dt}{z^3}  \nl
I_{bulk}^{5}  & = & - \frac{L}{4 \pi G} \int_{x_{min} }^{x_{int} } dx \int_{z_{E} }^{z_{RT_1} } dz \int_{0}^{t_{EW_1}} \frac{dt}{z^3} \nl
I_{bulk}^{6}  &=& - \frac{L}{4 \pi G} \int_{x_{int} }^{x_{max} } dx \int_{\varepsilon}^{z_{F} } dz \int_{0}^{t_{WDW} } \frac{dt}{z^3} \nl
I_{bulk}^{7} & = & - \frac{L}{4 \pi G} \int_{x_{int} }^{x_{max} } dx \int_{z_{F} }^{z_{RT_1} } dz \int_{0}^{t_{EW_1} } \frac{dt}{z^3} \, . 
\eea
All the integrals can be evaluated analytically. Since the expressions are rather 
cumbersome, we will write just the total expression of $\mathcal{C}_{V 2.0}^2$
 in (\ref{I-tot-bulk-disjoint-1}).

\subsection{Counterterms}

The counterterms for the null boundaries of the entanglement wedge vanish as usual.
We can separate the counterterm for the null boundaries of the WDW patch in two contributions:
\bea
 I_{ct,I} &=& \frac{L}{2 \pi G} \int_{x_{min} }^{x_{int} } dx \int_{\varepsilon}^{z_{G} } \frac{dz}{z^2} \, \log \le \frac{\tilde{L} \, \alpha \, z}{L^2} \ri  \, , \nl 
 I_{ct,II}&=&  \frac{L}{2 \pi G} \int_{x_{int} }^{x_{max} } dx \int_{\varepsilon}^{z_{F} } \frac{dz}{z^2} \, \log \le \frac{\tilde{L} \, \alpha \, z}{L^2} \ri \, .
\eea

\subsection{Joint contributions}

We have to include several joint contributions to the action:
\begin{itemize}
\item Joints on the cutoff at $z=\varepsilon$. The null normals are
\beq
\label{normals-wdw-disjoint}
\mathbf{k^{\pm}} = \alpha \le \pm dt - dz \ri \, ,
\eeq
and the contribution is:
\beq
I_{\varepsilon} = - \frac{L}{4 \pi G} \int_{x_{min} }^{x_{max}} dx \, \frac{\log \le \frac{\alpha^2 \varepsilon^2}{L^2} \ri }{\varepsilon} 
= - \frac{L}{2 \pi G} \frac{l \, \log \le \frac{\alpha \, \varepsilon}{L} \ri}{\varepsilon} \, .
\eeq

\item Joint on RT$_1$. The null normals to such surfaces are
\beq
\label{normals-wedge-big}
\mathbf{w_{1}^{\pm}} = \beta \le \pm dt + \frac{z}{\sqrt{z^2 + x^2}} dz + \frac{x}{\sqrt{z^2 + x^2}} dx \ri \, ,
\eeq
which gives
\beq
\begin{aligned}
I_{RT_1} & = - \frac{L}{2 \pi G} \int_{0}^{x_{max} } dx \, \frac{d+2 l}{\le d+ 2 l \ri^2 -4 x^2}
 \log  \frac{\beta^2  \left[ \le d+2 l \ri^2 - 4 x^2 \right]}{4 L^2} = \\
& = \frac{L}{4 \pi G} \log \le \varepsilon \ri \log \le \frac{\beta^2 \varepsilon}{L^2} \ri - \frac{L}{4 \pi G} 
\log \le d+ 2 l \ri \log \frac{\le d+2 l \ri \beta^2}{L^2} + \frac{L \pi}{48 G}  \, .
\end{aligned}
\eeq

\item Joint on RT$_2$. The null normals to these surfaces are
\beq
\label{normals-wedge-small}
\mathbf{w_{2}^{\pm}} = \gamma \le \pm dt - \frac{z}{\sqrt{z^2 + x^2}} dz - \frac{x}{\sqrt{z^2 + x^2}} dx \ri \, ,
\eeq
and the action is:
\beq
\begin{aligned}
I_{RT_2} & = - \frac{L}{2 \pi G} \int_{0}^{x_{min} } dx \, \frac{d}{d^ 2 -4 x^2} \log  \frac{\gamma^2  \le d^2 - 4 x^2 \ri}{4 L^2} = \\
& = \frac{L}{4 \pi G} \log \le \varepsilon \ri \log \le \frac{\gamma^2 \, \varepsilon}{L^2} \ri 
- \frac{L}{4 \pi G} \log \le d \ri \log \frac{d \, \gamma^2}{L^2} + \frac{L \pi}{48 G} \, .
\end{aligned}
\eeq

\item Joints between the two null boundaries of the entanglement wedge, curve $E$.
The normals are $ \mathbf{w_{1}^{+}} $ and $\mathbf{w_{2}^{+}}$.
The contribution gives 
\beq
I_{E}  = \frac{L}{\pi G} \int_{0}^{x_{int} } dx \, \frac{d+l}{\le d+l \ri^2 -4 x^2} \log \left[ \frac{\beta \, \gamma \left[ \le d+l \ri^2 -4 x^2 \right]}{4 L^2} \right] \, .
\eeq

\item Joint between  the null boundary of the WDW patch and the null boundary 
of the entanglement wedge anchored at RT$_1$, curve $F$.
The  normals are $\mathbf{k^{+}}$ and $\mathbf{w_{1}^{+}} $.
The term gives
\beq
I_{F}  = \frac{2 L}{\pi G} \int_{x_{int} }^{x_{max} } dx \, \frac{d+ 2 \le l + \varepsilon \ri}{\le d+ 2 \le l + \varepsilon \ri \ri^2 - 4 x^2} \log \left[ \frac{\alpha \, \beta \left[ \le d + 2 \le l + \varepsilon \ri \ri^2 - 4 x^2 \right]^2}{16 L^2 \left[ \le d + 2 \le l + \varepsilon \ri \ri^2 + 4 x^2 \right]} \right] \, .
\eeq

\item Joint between
 the null boundary of the WDW patch and the null boundary of the entanglement wedge anchored at RT$_2$ (curve $G$)
 with normals $\mathbf{k^{+}} $ and $ \mathbf{w_{2}^{+}}$.
 The contribution gives
\beq
I_{G}  = \frac{2 L}{\pi G} \int_{x_{min} }^{x_{int} } dx \, \frac{d-2 \varepsilon}{4 x^2 - \le d - 2 \varepsilon \ri^2 } \log \left[ \frac{\alpha \, \gamma \le d -2 \varepsilon +2 x \ri^2 \le d-2 \varepsilon -2 x \ri^2}{16 L^2 \left[ 4 x^2 + \le d-2 \varepsilon \ri^2 \right]} \right] \, .
\eeq

\end{itemize}

\subsection{Complexities}
Adding up all the contributions and using polylog identities, we find:
\beq
\label{CA2}
\begin{aligned}
\mathcal{C}_A^2 & =  \frac{c}{3\pi^2 }  \left\{ \log \le \frac{\tilde{L}}{L} \ri \frac{l}{\varepsilon} 
 -  \log \le \frac{2 \tilde{L}}{L} \ri \log \le \frac{d(d+2l)}{\varepsilon^2} \ri - \frac{\pi^2}{4}  \right.  \\
& + \left[  \log \le \frac{\tilde{L}}{L} \ri 
+ \log \le \frac{2(d+l)}{\sqrt{d(d+2l)}} \ri \right]
 \log \le \frac{(d+l+\sqrt{d(d+2l)})^2}{l^2} \ri  \\
& \left. + \mathrm{Li}_{2} \le \frac{\sqrt{d (d+2l)}}{d+l} \ri - \mathrm{Li}_{2} \le - \frac{\sqrt{d (d+2l)}}{d+l} \ri \right\} \, .
\end{aligned}
\eeq
The spacetime volume complexity instead is
\bea
\label{I-tot-bulk-disjoint-1}
\mathcal{C}_{V 2.0}^2 &=& \frac{2 \, c}{3} \left\{ \frac{2 l}{\varepsilon} - 2 \log \frac{d(d+2l)}{\varepsilon^2}
 + \frac{ \pi^2}{2 }   + 8 \, \mathrm{arctanh} \, \sqrt{\frac{d}{d+2l}}  \right. \nl
&& \,\,\,\,\,\, \left.
- 2  \left[ {\rm Li}_{2} \le \frac{\sqrt{d (d+2l)}}{d+l} \ri - {\rm Li}_{2} \le - \frac{\sqrt{d (d+2l)}}{d+l} \ri \right]  \right\} \, .
\eea
The divergences of (\ref{CA2})  and (\ref{I-tot-bulk-disjoint-1})
are respectively  the same as in eqs. (\ref{disjoint-complexity-case1});
in particular, the subleading divergences are still poportional to the entanglement entropy
\beq
S = \frac{c}{3} \log \frac{d(d+2l)}{\varepsilon^2} \, .
\eeq
The  finite part instead is a more complicated function of $d,l$
 compared to the single interval case.

\section{ Mutual complexity}
\label{sec:mutual}

Consider a physical system which is splitted into two sets $A,B$.
The mutual information is defined as
\beq
I(A|B) = S (A) + S (B) - S (A \cup B) \, .
\eeq
Since the entanglement entropy is shown to exhibit a subadditivity behaviour, 
\emph{i.e.} the entanglement entropy of the full system is less than the sum of the entropies
 related to the two subsystems, the mutual information is a positive quantity.

Another quantity which measures the correlations between
 two physical subsystems was defined in \cite{Alishahiha:2018lfv,Caceres:2019pgf}
  and called \emph{mutual complexity}:
\beq
\Delta \mathcal{C}  =
\mathcal{C}(\hat{\rho}_A)+\mathcal{C}(\hat{\rho}_B) - \mathcal{C}(\hat{\rho}_{A \cup B}) \, .
\label{mutual1}
\eeq
where $\hat{\rho}_A$, $\hat{\rho}_B$ are the reduced density matrices 
in the Hilbert spaces localised in  $A$ and $B$.
If $\Delta \mathcal{C} $ is always positive, complexity is subadditive;
if it is always negative, complexity  is superadditive.
By construction, in the CV and CV 2.0 conjectures complexity is always 
superadditive, i.e.  $\Delta \mathcal{C} \leq 0$. Instead, in the $CA$ conjecture,
 no general argument is known which fixes the sign of $\Delta \mathcal{C}$.

$\Delta \mathcal{C}$ is a finite quantity in all the three holographic
conjectures. Moreover,  $\Delta \mathcal{C} =0$ for $d>d_0$
 because in this case  the RT surface is  disconnected and then 
$\mathcal{C}(\hat{\rho}_A)+\mathcal{C}(\hat{\rho}_B) = \mathcal{C}(\hat{\rho}_{A \cup B}) $.
We will check that this quantity  is generically discontinuous
at $d=d_0$.  

\begin{figure}[h]
\center
\begin{tabular}{ccc}
\includegraphics[scale=0.5]{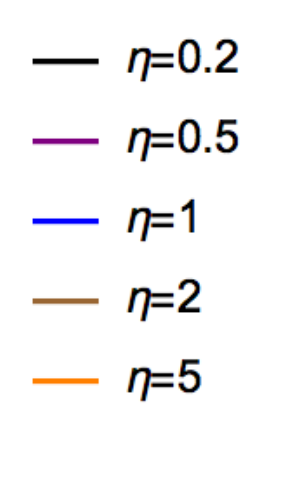} &
\includegraphics[scale=0.5]{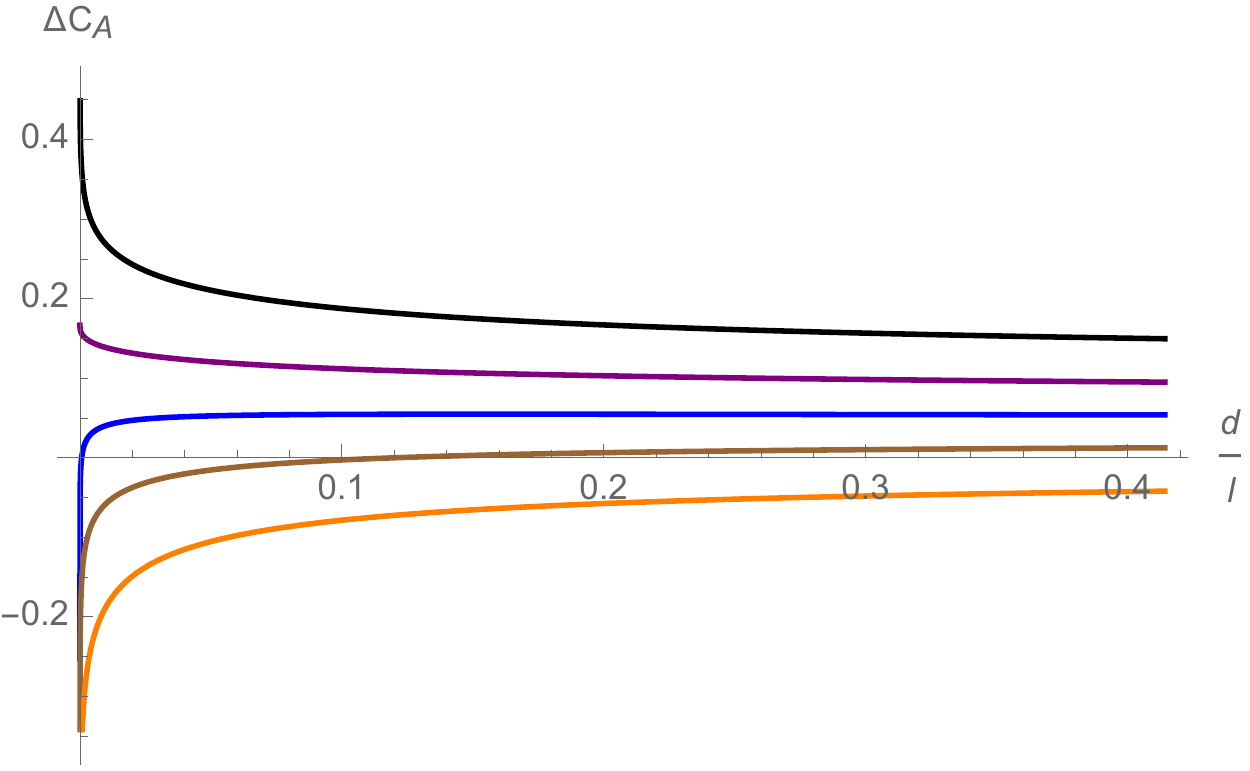} & 
 \includegraphics[scale=0.5]{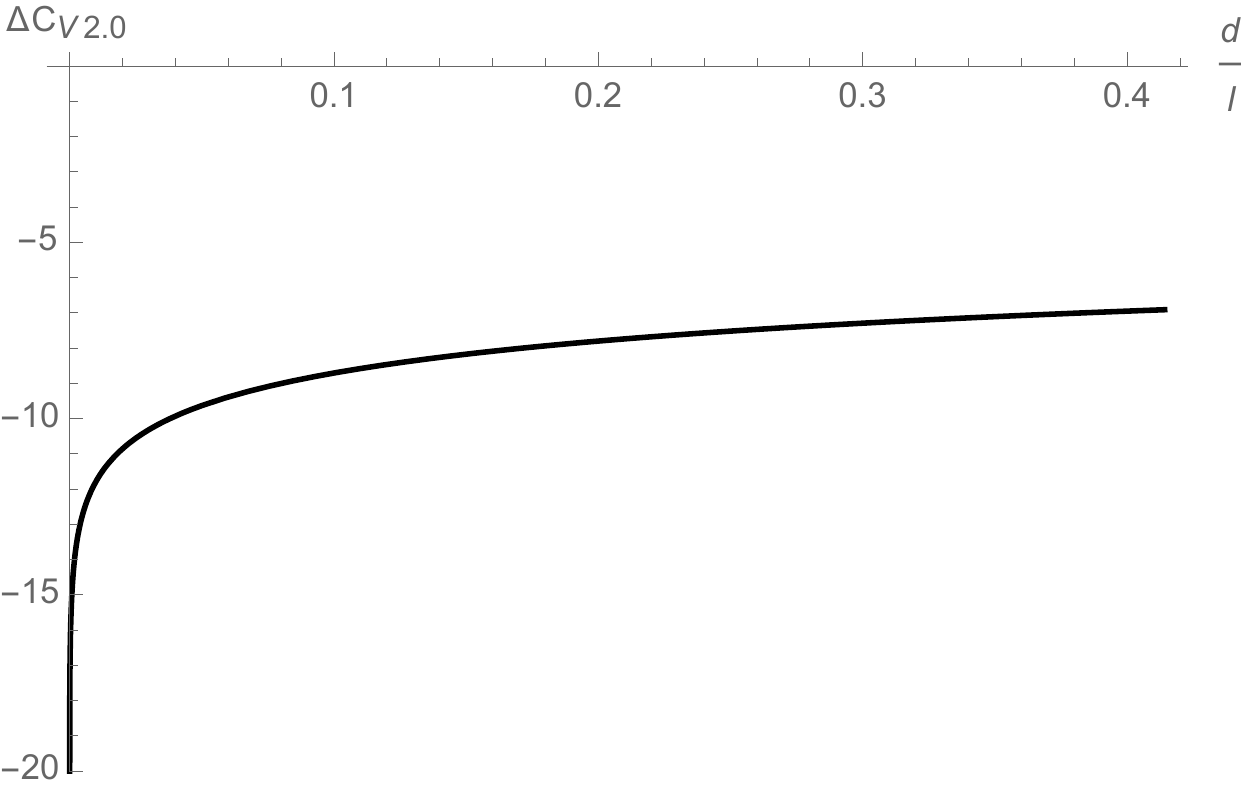}
\end{tabular}
\caption{Left: Mutual complexity $\Delta \mathcal{C}_A$ for
 several values of $\eta=\tilde{L}/L$  as a function of $\frac{d}{l} \in [0,\frac{d_0}{l}=\sqrt{2}-1]$. 
 In order to have a positive-definite complexity, we must impose $\eta>1$. The other values of $\eta$
 have been included for illustrative purpose.
Right: Mutual complexity $\Delta \mathcal{C}_{V 2.0}$.
Here we set $c=1$.}
\label{mutual}
\end{figure}

In the case of two disjoint intervals, from eq. (\ref{CA2}) we find that
the action mutual complexity is:
\beq
\begin{aligned}
\Delta \mathcal{C}_A & =  \mathcal{C}_A^1 - \mathcal{C}_A^2 
 =  \frac{c}{3 \pi^2 } \left\{  \log \le \frac{2 \tilde{L}}{L} \ri \log \le \frac{d(d+2l)}{l^2} \ri + \frac{\pi^2}{2} \right.   \\
& 
- \left[  \log \le \frac{\tilde{L}}{L} \ri 
+ \log \le \frac{2(d+l)}{\sqrt{d(d+2l)}} \ri \right]
 \log \le \frac{(d+l+\sqrt{d(d+2l)})^2}{l^2} \ri  \\
& \left. - \mathrm{Li}_{2} \le \frac{\sqrt{d (d+2l)}}{d+l} \ri + \mathrm{Li}_{2} \le - \frac{\sqrt{d (d+2l)}}{d+l} \ri \right\} \, .
\end{aligned}
\eeq
The function $\Delta \mathcal{C}_A$ 
is plotted in figure \ref{mutual} for various $\eta=\tilde{L}/L$. 
From the figure, we see that this quantity can be either positive or negative.
At small $d$, the behavior of  $\Delta \mathcal{C}_A$ is:
\beq
\Delta \mathcal{C}_A \approx \frac{c}{3 \pi^2} 
\log \le \frac{2 \tilde{L}}{L} \ri \log \le  \frac{2 d}{l} \ri \, .
\eeq
For the value $\tilde{L}/L=1/2$, the behaviour of $\Delta \mathcal{C}_A$ at $d \rightarrow 0$
switches from $-\infty$ to $\infty$.

If $\eta \leq 1/2$, CA is subadditive for all values of $d/l$.
For $\eta > 1/2, $ it is always possible to find small enough distances giving a superadditive behaviour.
Moreover, there is a critical $\eta_0 \approx 2.465$ in such a way that complexity 
of two disjoint  intervals is always superadditive if $\eta>\eta_0$.
In order to have a positive definite subregion complexity,
we should require $\eta>1$. Then it is not possible to achieve
an universally subadditive complexity in a physically consistent setting.

A similar behaviour of subregion CA is found
in the thermofield double state where the subsystems are taken as
 the two disconnected boundaries of spacetime.
This case was investigated for asymptotically $\mathrm{AdS}$ black holes in $D$ dimensions 
\cite{Agon:2018zso,Alishahiha:2018lfv}, 
showing that the complexity=action is subadditive when $\eta<\hat{\eta}_D$ and
superadditive for $\eta>\hat{\eta}_D$.
The value of $\hat{\eta}_D$ is given by the zero of $g_D(\eta)$ \cite{Agon:2018zso}:
\beq
g_D(\eta)= \log( (D-2) \eta)+\frac12 \le \psi_0(1)-\psi_0\le \frac{1}{D-1} \ri \ri +\frac{D-2}{D-1} \pi \, ,
\eeq
where $\psi_0(z)=\Gamma'(z)/\Gamma(z)$ is the digamma function.
For $D=3$,  $\hat{\eta}_3 \approx 0.1$. 

In the CV 2.0 conjecture,  from eq. (\ref{I-tot-bulk-disjoint-1}) we find that
the mutual complexity for two disjoint intervals is:
\beq
\begin{aligned}
\Delta \mathcal{C}_{V 2.0} & =  \mathcal{C}_{V 2.0}^1 - \mathcal{C}_{V 2.0}^2 =   \frac{4 \, c}{3} \left[
 \log \frac{d(d+2l)}{ l^2} - \frac{ \pi^2}{2}  - 4 \, \mathrm{arctanh} \, \sqrt{\frac{d}{d+2l}} \right. \\
 &  \left. +   {\rm Li}_{2} \le \frac{\sqrt{d (d+2l)}}{d+l} \ri - {\rm Li}_{2} \le - \frac{\sqrt{d (d+2l)}}{d+l} \ri \right] \, ,
\end{aligned}
\eeq
see figure \ref{mutual} for a plot.
This is negative definite as expected, because the bulk region 
involved in the first configuration of RT surface is smaller than the second region.

In the CV conjecture, we can use eqs.  (\ref{CCVA}) and  (\ref{CCVB}) from \cite{Abt:2017pmf} 
to determine mutual complexity.
Considering the case of a double segment,
we find that for $d<d_0$ the mutual complexity is constant:
\beq
\Delta \mathcal{C}_{V}= -\frac{4 \, c}{3} \pi \, .
\eeq

\subsection{Strong super/subaddivity for overlapping segments}

Given two generically overlapping regions $A$ and $B$,
entanglement entropy satisfies the strong subadditivity property:
\beq
\tilde{\Delta} S = S (A) + S(B) -S(A \cup B) - S(A \cap B) \geq 0 \, .
\label{gene}
\eeq
Inspired by this relation, we can define \cite{Caceres:2019pgf}
by analogy a generalization of the mutual complexity as:
\beq
\tilde{\Delta} \mathcal{C}(A,B) =
\mathcal{C}(\hat{\rho}_A)+ \mathcal{C}(\hat{\rho}_B) - \mathcal{C}(\hat{\rho}_{A \cup B}) - \mathcal{C}(\hat{\rho}_{A \cap B}) \, .
\eeq
This definition generalizes eq. (\ref{mutual1}) to the case where $A \cap B \ne \emptyset$.
We can investigate the sign of this quantity in the case of two overlapping segments.
 
 Suppose that we consider the regions given by 
 two intervals of lengths $a,b$ which intersect 
in a segment of length $c$. The union of these intervals 
is a segment of total length $a+b-c$. From eqs. (\ref{ACTION-C}) and (\ref{total-btz-bulk}), we find
\bea
\tilde{\Delta} \mathcal{C}_A^{\rm BTZ}&=&-\log \le \frac{2 \tilde{L}}{L}\ri \tilde{\Delta} S^{\rm BTZ} \, , \nl
\tilde{\Delta} \mathcal{C}_{V 2.0}^{\rm BTZ}&=&-4 \tilde{\Delta} S^{\rm BTZ} \, ,
\eea
where $\tilde{\Delta} S^{\rm BTZ}$ is the quantity defined in (\ref{gene}), computed
 for the two overlapping intervals in the BTZ background.
Then $ \mathcal{C}_A$ is strongly subadditive for $\tilde{L}/L < 1/2$
and strongly superadditive for $\tilde{L}/L > 1/2$. 
Instead $ \mathcal{C}_{V 2.0}$ is strongly superadditive.


\section{Conclusions}

We studied the CA and CV 2.0 subregion complexity
conjectures in AdS$_3$ and in the BTZ background.
The main results of this paper are:
\begin{itemize}
\item
In the case of one segment, we find that subregion complexity for AdS$_3$ and for the BTZ
 can be directly related to the entanglement entropy, see eqs. (\ref{CABTZ}) and  (\ref{total-btz-bulk}).
\item
 In the case of a two segments subregion,
complexity in AdS$_3$ is a more complicated function of 
the lengths and the relative separation of the segments, see eqs. (\ref{CA2}) and (\ref{I-tot-bulk-disjoint-1}).
Subregion complexity carries a 
different  amount of information compared to the entanglement entropy. In particular,
for two disjoint segments the mutual complexity (defined in eq. (\ref{mutual1})) is not
proportional to mutual information. 
\end{itemize}

One of the obscure aspects of the CA conjecture is the physical meaning
of the scale $\tilde{L}$ appearing in the action counterterm eq. (\ref{counterterm})
on the null boundaries. A deeper understanding of the role of this parameter
is desirable. In particular, its relation with the field theory side of the correspondence
remains completely unclear.

For $d<d_0=(\sqrt{2}-1) l$,
we find that the sign of action mutual complexity $\Delta \mathcal{C}_A$
 of a two disjoint segments subregion
depends drastically on $\eta=\tilde{L}/L$ (see figure \ref{mutual})\footnote{For $d>d_0$, $\Delta \mathcal{C}_A$ instead vanishes for all 
the values of $\eta$.}.
First of all, we have to impose $\eta>1$ in order to have a positive-definite
complexity. Then there are two different cases:
\begin{itemize}
\item $ \eta \geq \eta_0 \approx 2.465 $, $\Delta \mathcal{C}_A$ is always negative, and so
$\mathcal{C}_A$ is superadditive as $\mathcal{C}_V$ and  $\mathcal{C}_{V 2.0}$. 
\item $1 < \eta < \eta_0 $,  $\Delta \mathcal{C}_A$ is negative at small 
$d$ and positive at larger $d<d_0$.  In this interval,  $\mathcal{C}_A$ 
is superadditive only at small enough $d$, and it is  subadditive at larger $d$.
\end{itemize}
It would be interesting to study the case of higher dimensional AdS space, in order
to investigate the behaviour of mutual complexity for regions of different shape,
such as a higher dimensional spheres or  strips.

Sub/superadditivity properties of complexity in quantum systems
should depend on the specific definition of subregion complexity
which is adopted, e.g. purification or basis complexity as defined in \cite{Agon:2018zso}.
So far, all the studies we are aware of focused on purification complexity,
see e.g. \cite{Camargo:2018eof,Caceres:2019pgf}. The situation looks
rather intricated: it seems that both sub and superadditive behaviours 
might be obtained, depending on the explicit basis used to define complexity
and other details (see for example table 1 of \cite{Caceres:2019pgf}).

In the CV conjecture, subregion complexity for multiple intervals
 in the BTZ background is independent of
temperature and can be computed using topology from the Gauss-Bonnet theorem,
 see  \cite{Abt:2017pmf}. It would be interesting to investigate if a similar
 relation with topology holds also for CA and CV 2.0.
 The complicated structure of the finite terms in eqs.
 (\ref{CA2}) and (\ref{I-tot-bulk-disjoint-1}) suggests that 
 such relation, if exists, is more intricated than in CV.


 \section*{Acknowledgments}
One of the authors (PR) thanks the Dipartimento di Matematica e Fisica, Universit\`a Cattolica del Sacro Cuore, 
Brescia, Italy for supporting a visit during which this work was initiated.


\section*{Appendix}
\addtocontents{toc}{\protect\setcounter{tocdepth}{1}}
\appendix

\section{Another regularization for the action of one segment in BTZ}
\label{app:other-reg}

\begin{figure}[h]
\center
\begin{tabular}{cc}
\includegraphics[scale=0.5]{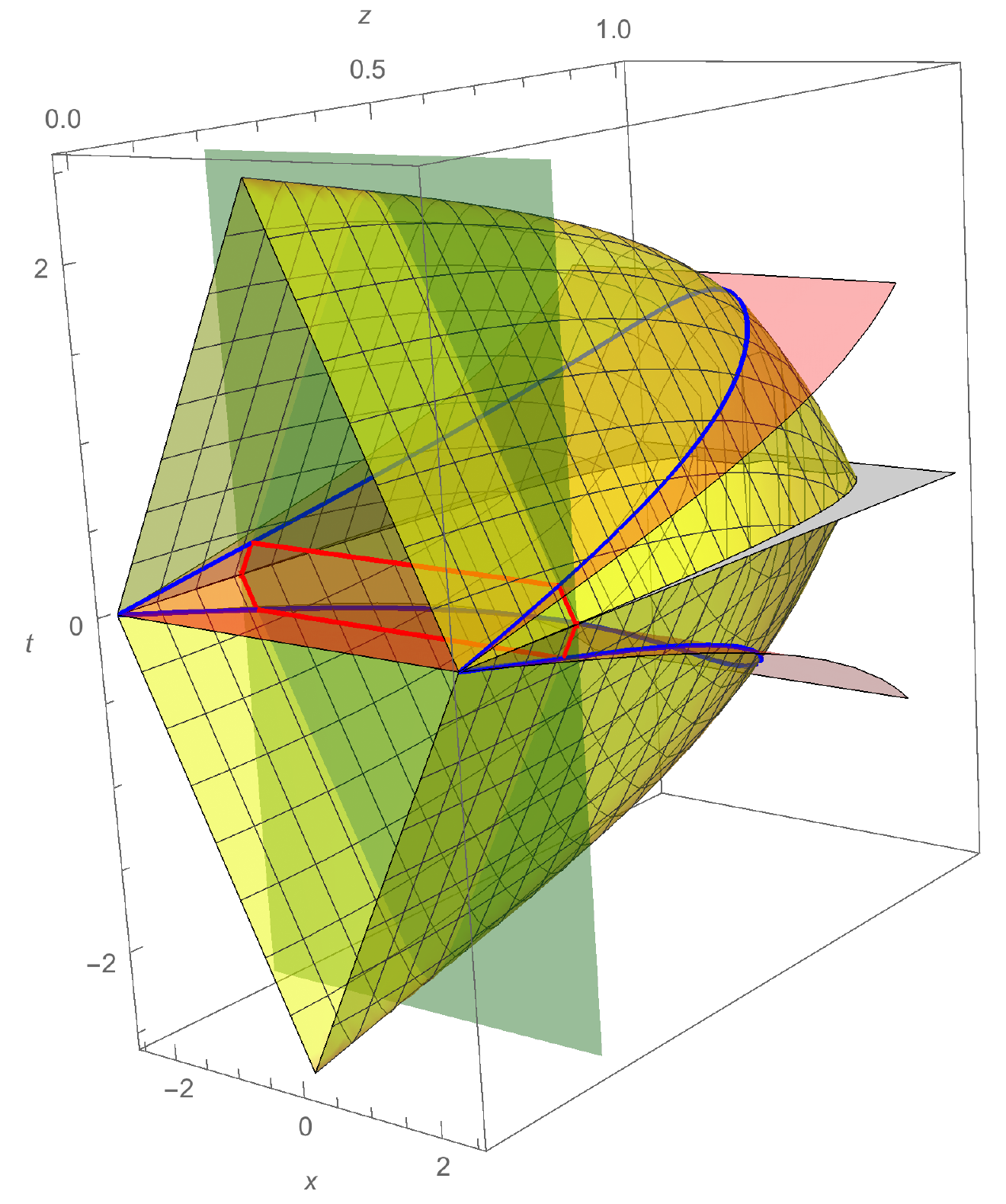} & \includegraphics[scale=0.55]{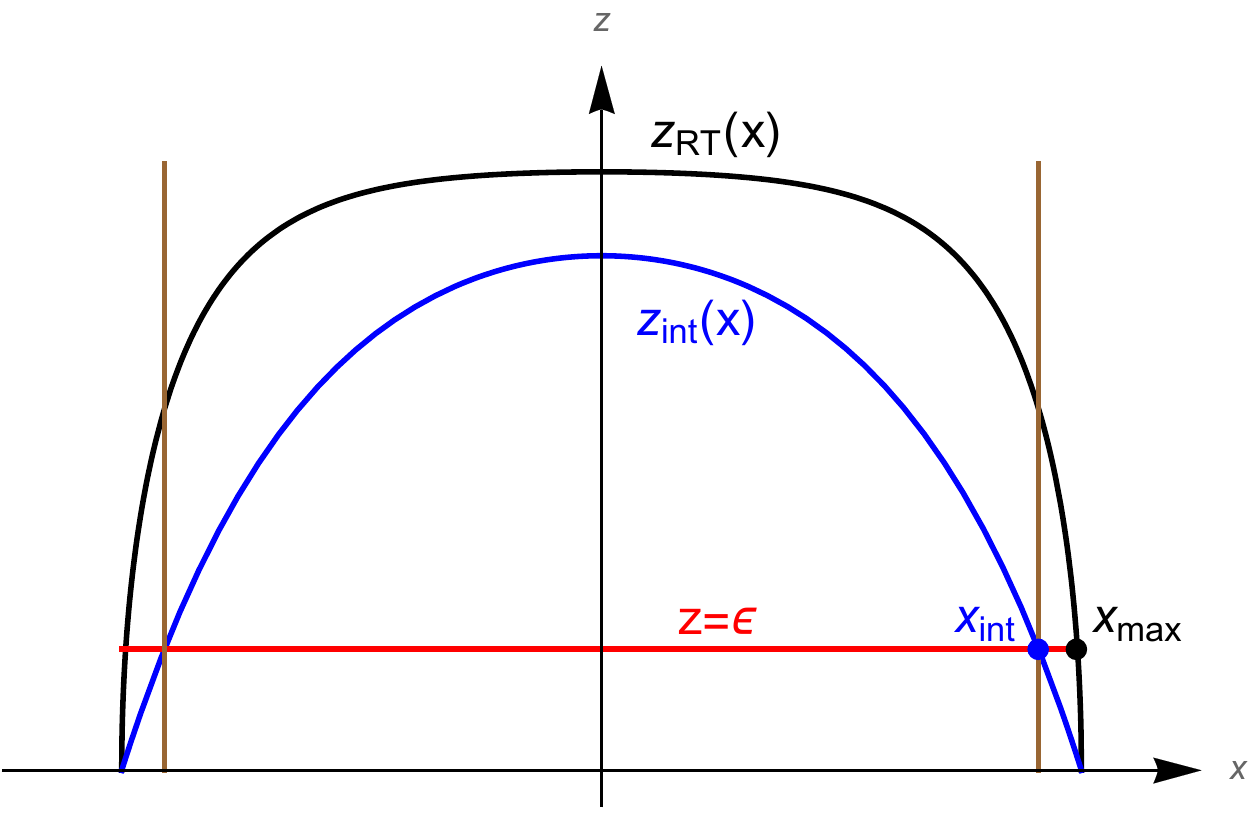}
\end{tabular}
\caption{ Another regularization for the BTZ case.}
\label{fig-BTZ-another}
\end{figure}

 In this Appendix we follow another prescription to regularize
  the action where the null boundaries of the WDW patch are sent from
   the true boundary $z=0$ and we add a timelike cutoff surface
    at $z=\varepsilon$ cutting the bulk structure we integrate over. 
   The geometry of the region is shown in figure \ref{fig-BTZ-another}.

The geometric data are slightly different than the ones introduced in Section \ref{sect:BTZ}.
The RT surface and the corresponding entanglement wedge are the same, see eqs. (\ref{RT surface}) and (\ref{twedge}).
The WDW patch starts from the true boundary $z=0$ and then the null lines which delimit it are parametrized by
\beq
t_{\rm WDW} = \pm \frac{z_h}{4} \log \le \frac{z_h + z}{z_h - z} \ri^2 \, ,
\eeq
where $\pm$ refers to positive and negative times, respectively.
The intersection curve between the WDW patch and the entanglement wedge is given by 
\beq
z_{\rm int} = \coth \le \frac{l}{2 z_h} \ri - \cosh \le \frac{x}{z_h} \ri  \mathrm{csch}  \le \frac{l}{2 z_h} \ri \, .
\label{zint 1st regularization}
\eeq
The null normals to the boundaries of the WDW patch and the entanglement wedge are unchanged.

Unlike the case of the other regularization, the intersection curve and the
 RT surface do not meet at $z=\varepsilon,$ but at the true boundary $z=0.$
For this region, there are no codimension-3 joints. 
The intersection curve between the boundaries of the WDW patch and the entanglement wedge
meets the cutoff surface at:
\beq
x_{\rm int}= \mathrm{arccosh} \, \left[ \cosh \le \frac{l}{2 z_h} \ri - \frac{\varepsilon}{z_h} \sinh  \le \frac{l}{2 z_h} \ri \right] \, .
\eeq
This expression is found by inverting eq. (\ref{zint 1st regularization}) and imposing $z=\varepsilon.$
In the folowing sections we compute all the terms entering the gravitational action.

\subsection{Bulk contribution}
\label{Bulk contribution 1st regularization}

We split the contributions as follows
\begin{equation}
I_{\rm bulk} = 4 \left( I_{\rm bulk}^{1} + I_{\rm bulk}^{2} + I_{\rm bulk}^{3} \right) \, ,
\end{equation}
where
\bea
I_{\rm bulk}^{1} &=& - \frac{L}{4 \pi G} \int_{0}^{x_{\rm int} } dx \int_{\varepsilon}^{z_{\rm int}} dz \int_{0}^{t_{\rm WDW}} dt \, \frac{1}{z^{3}} \, ,
\nl
I_{\rm bulk}^{2} &=& -  \frac{L}{4 \pi G} \int_{0}^{x_{\rm int} } dx \int_{z_{\rm int}}^{z_{\rm RT}} dz \int_{0}^{t_{\rm EW}} dt \, \frac{1}{z^{3}} \, ,
\nl
I_{\rm bulk}^{3} &=& -  \frac{L}{4 \pi G} \int_{x_{\rm int}}^{{x_{\rm max}} } dx \int_{\varepsilon}^{z_{\rm RT}} dz \int_{0}^{t_{\rm EW}} dt \, \frac{1}{z^{3}} \, .
\eea
In this case the sum of bulk terms obtained by
 splitting the spacetime region with the intersection between the boundaries of the WDW 
 patch and the entanglement wedge does not give the entire bulk action.
We need to add  $I_{\rm bulk}^{3}$ which accounts 
for the region between the values $x_{int}$ and $x_{max}$ of the transverse coordinate.

A direct evaluation gives
\beq
\begin{aligned}
 I^1_{\rm bulk}  + I_{\rm bulk}^{2}  = \frac{L}{16 \pi G z_h} & \int_{0}^{x_{\rm int}(\varepsilon)} dx \, \left\lbrace \coth \le \frac{x}{z_h} \ri \log \left| \frac{\sinh \le \frac{l-2x}{2 z_h} \ri  \sinh^2 \left[ \frac{l+2x}{4 z_h} \right]}{\sinh \le \frac{l+2x}{2 z_h} \ri  \sinh^2 \left[ \frac{l-2x}{4 z_h} \right]}  \right| \right. \\
& \left.  +  \frac{2 \sinh \le \frac{l}{2 z_h} \ri}{\cosh \le \frac{l}{2 z_h} - \cosh \le \frac{x}{z_h} \ri \ri} - \frac{2 z_h}{\varepsilon}  + \le \frac{z_h^2}{\varepsilon^2} - 1 \ri \log \left| \frac{z_h - \varepsilon}{z_h + \varepsilon}  \right|
   \right\rbrace \, .
 \end{aligned}
\eeq
\beq
I^{3}_{\rm bulk}   = - \frac{L}{16 \pi G} \, .
\label{bulk 3 1st reg}
\eeq

\subsection{Gibbons-Hawking-York contribution}
The Gibbons-Hawking-York (GHY) surface term in the action for timelike and spacelike boundaries is
\begin{equation}
I_{GHY} = \frac{1}{8 \pi G} \int_{\partial \mathcal{B}'} d^{2} x \sqrt{- \det  h_{\mu \nu}} \, K
\end{equation}
with $h_{\mu \nu}$ the induced metric on the boundary and $K$ the trace of the extrinsic curvature. 
The only contribution of this kind comes from the timelike regularizing surface at $z = \varepsilon$.

The GHY contribution is given by two parts. The first one involves the WDW patch, while the second one involves the entanglement wedge:
\begin{equation}
I_{\rm GHY}^{1} = \left[ \frac{L}{8 \pi G}  \int_{0}^{x_{\rm int} } dx \int_{0}^{t_{\rm WDW} } dt \, \le  \frac{2}{z^2} - \frac{1}{z_h^2} \ri \right]_{z = \varepsilon} = \frac{L}{8 \pi G} \, \frac{l}{\varepsilon} - \frac{L}{4 \pi G}  \, ,
\label{GHY sotto xint 1st reg}
\end{equation}
\begin{equation}
I_{\rm GHY}^{2} = \left[ \frac{L}{8 \pi G}  \int_{x_{\rm int} }^{x_{max} } dx \int_{0}^{t_{\rm EW} } dt \,  \le  \frac{2}{z^2} - \frac{1}{z_h^2} \ri \right]_{z = \varepsilon} = \frac{L}{8 \pi G} \, .
\label{GHY sopra xint 1st reg}
\end{equation}
The total GHY contribution is
\begin{equation}
\label{ghy1}
I_{\rm GHY} = 4 \left( I_{\rm GHY}^{1} + I_{\rm GHY}^{2} \right) = \frac{L}{2 \pi G} \left( \frac{l}{\varepsilon} - 1 \right) \, .
\end{equation}

\subsection{Null boundaries counterterms}

The details of calculation are very similar to the ones in section \ref{sect-BTZ-countertems}.
The contribution in eq. (\ref{null-bou}) and  the counterterm on the boundary of entanglement
wedge again vanish. The counterterm on the boundary of the WDW patch gives:
\beq
\begin{aligned}
I_{\rm ct}^{\rm WDW} & = - \frac{L}{2 \pi G} \int_0^{x_{\rm int} } dx \int_{\varepsilon}^{z_{\rm int} } dz \, \frac{1}{z^2} \log \left| \frac{\tilde{L}}{L^2} \, \alpha z \right| = \\
& =  \frac{L}{2 \pi G} \int_0^{x_{\rm max}} dx  \, \left\lbrace \frac{1 + \log \left| \frac{\tilde{L}}{L^2} \, \alpha \varepsilon \right|}{\varepsilon} + \frac{\sinh \le \frac{l}{2 z_h}  \ri}{z_h \left[ \cosh \le \frac{x}{z_h} \ri - \cosh \le \frac{l}{2 z_h} \ri   \right]}  \times \right. \\
& \left. \times \le 1 + \log \left| \frac{\tilde{L} z_h \alpha}{L^2} \frac{\cosh \le \frac{l}{2 z_h} \ri - \cosh \le \frac{x}{z_h} \ri}{\cosh \le \frac{l}{2 z_h}  \ri} \right|  \ri  \right\rbrace \, .
\end{aligned}
\eeq

\subsection{Joint terms}

The joint contribution to the gravitational action coming from a codimension-$2$ 
surface given by the intersection of a codimension-$1$ null surface and a codimension-$1$ timelike (or spacelike) surface is 
\begin{equation}
\label{null-time joint}
I_{\mathcal{J}} = \frac{\eta}{8 \pi G} \int_{\mathcal{J}} dx \, \sqrt{\sigma} \log \left| \mathbf{k} \cdot \mathbf{n} \right| \, ,
\end{equation}
where $\sigma$ is the induced metric determinant on the codimension-$2$ surface and $\mathbf{n}$ and $\mathbf{k}$ are the outward-directed normals to the timelike (or spacelike) surface and the null one respectively. Moreover,
\begin{equation}
\label{sign null-time}
\eta = - \, \mathrm{sign} \left( \mathbf{k} \cdot \mathbf{n} \right) \, \mathrm{sign} \left( \mathbf{k} \cdot \hat{t} \right) 
\end{equation} 
in which $\hat{t}$ is the auxiliary unit vector in the tangent space of the boundary region, 
orthogonal to the joint and outward-directed from the region of interest \cite{Carmi:2016wjl}.

The unit normal vector $n^{\mu}$ to the $z = \varepsilon $ surface is
\begin{equation}
\label{nvector}
n^{\mu} = \left( 0, \, - \frac{z}{L} \sqrt{f(z)}, \, 0 \right) 
\end{equation}
where the sign must be chosen so that the vector is outward-directed from the region of interest. 

The joints give the following contributions:
\begin{itemize}
\item
The joint involving the WDW patch boundary and the cutoff surface:
\begin{equation}
I_{\mathcal{J}}^{\rm cutoff1} = - \, \frac{L}{2 \pi G} \int_{0}^{x_{\rm int} } \frac{dx}{\varepsilon} \log \left( \frac{\alpha \, \varepsilon}{L \sqrt{f(\varepsilon)}} \right) =
- \frac{L}{4 \pi G} \frac{l}{\varepsilon} \log \le \frac{L}{\alpha \varepsilon} \ri - \frac{L}{2 \pi G} \log \le \frac{L}{\alpha \varepsilon} \ri \, .
\label{joint cutoff 1 sotto xint 1st reg}
\end{equation}
\item
Next we consider the joint involving the cutoff surface and the entanglement wedge boundary:
\begin{equation}
I_{\mathcal{J}}^{\rm cutoff2} = \mathcal{O} \left( \varepsilon \log \varepsilon \right) \, .
\label{joint cutoff 1 sopra xint 1st reg}
\end{equation}
\item The null-null joint contribution coming from the RT surface 
is  the same as in the previous regularization, see eq. (\ref{joint-BTZ-RT}).
\item
The joints coming from the intersection between the null boundaries of the 
WDW patch and the ones of the entanglement wedge 
give a similar contribution  as in eq. (\ref{joint-BTZ-EW}),
The main difference is that the integral is in  the range $ [0, x_{\rm int} (\varepsilon)] $ 
and the intersection is slightly different, because the WDW patch starts from $z=0$ in the present regularization:
\begin{equation}
I_{\mathcal{J}}^{\rm int} =  \frac{L}{2 \pi G z_h } \int_{0}^{x_{\rm int }} dx \, 
\frac{\sinh \le  \frac{l}{2 z_h} \ri}{\cosh \le \frac{l}{2 z_h} \ri - \cosh \le \frac{x}{z_h} \ri } \,
\log \left| \frac{\alpha \beta z_h^2}{2 L^2} \frac{\left(\cosh \le \frac{l}{2 z_h} \ri - \cosh \le \frac{x}{z_h} \ri \right)^2}{\cosh \le \frac{x}{z_h} \ri \cosh \le \frac{l}{2 z_h} \ri -1}  \right| \, .
\end{equation}
\end{itemize}

\subsection{Complexity}
Adding all the contributions and performing the integrals we finally get
\beq
 \mathcal{C}_{A}^{\rm BTZ} = \frac{l}{\varepsilon} \frac{c}{6 \pi^2} \le 1 +\log \left(\frac{\tilde{L}}{L} \right) \ri
- \log  \left(\frac{2\tilde{L}}{L} \right) \frac{S^{\rm BTZ}}{\pi^2} - \frac{c}{3 \pi^2} \le \frac12 + \log \le \frac{\tilde{L}}{L} \ri \ri + \frac{1}{24} c \, .
\label{ACTION-C 1st reg}
\eeq
The difference with expression (\ref{ACTION-C}) consists 
only in the coefficient of the divergence $1/\varepsilon$ 
and in a finite piece proportional to the counterterm scale $\tilde{L}$ via a logarithm.

Recently other counterterms were proposed to give a universal behaviour
 of all the divergences of the action \cite{Akhavan:2019zax}.
In particular, with this regularization we need to insert a codimension-1 
boundary term at the cutoff surface:
\beq
I_{\rm ct}^{\rm cutoff} = - \frac{1}{16 \pi G} \int d^{d-1} x \, dt \, \sqrt{-h} \, 
\le \frac{2(d-1)}{L} + \frac{L}{d-2} \tilde{R} \ri  \, ,
\label{akhavan}
\eeq
being $\tilde{R}$ the Ricci scalar on the codimension-1 surface.
Adding the extra counterterm in eq. (\ref{akhavan}), we find
\beq
  \mathcal{C}_{A}^{\rm BTZ} = \frac{l}{\varepsilon} \frac{c}{6 \pi^2} \log \left(\frac{\tilde{L}}{L} \right) 
- \log  \left(\frac{2\tilde{L}}{L} \right) \frac{S^{\rm BTZ}}{\pi^2} - \frac{c}{3 \pi^2} \log \le \frac{\tilde{L}}{L} \ri  + \frac{1}{24} c \, .
\label{ACTION-C 1st reg plu new counterterm}
\eeq
The numerical coefficient of all the divergences is the same as in eq. (\ref{ACTION-C}).
The two regularizations differ only by a finite piece dependent
 from the counterterm length scale $\tilde{L}.$

\end{document}